\documentclass[%
 reprint,
superscriptaddress,
nofootinbib,
 amsmath,amssymb,
 aps
]{revtex4-2}

\usepackage{graphicx}% Include figure files
\usepackage{dcolumn}% Align table columns on decimal point
\usepackage{bm}% bold math
\usepackage{aasmacros}
\usepackage{float}
\usepackage{soul, xcolor}
\usepackage{ulem}
\usepackage{scalerel}
\usepackage{tikz}
\usetikzlibrary{svg.path}

\definecolor{orcidlogocol}{HTML}{A6CE39}
\tikzset{
  orcidlogo/.pic={
    \fill[orcidlogocol] svg{M256,128c0,70.7-57.3,128-128,128C57.3,256,0,198.7,0,128C0,57.3,57.3,0,128,0C198.7,0,256,57.3,256,128z};
    \fill[white] svg{M86.3,186.2H70.9V79.1h15.4v48.4V186.2z}
                 svg{M108.9,79.1h41.6c39.6,0,57,28.3,57,53.6c0,27.5-21.5,53.6-56.8,53.6h-41.8V79.1z M124.3,172.4h24.5c34.9,0,42.9-26.5,42.9-39.7c0-21.5-13.7-39.7-43.7-39.7h-23.7V172.4z}
                 svg{M88.7,56.8c0,5.5-4.5,10.1-10.1,10.1c-5.6,0-10.1-4.6-10.1-10.1c0-5.6,4.5-10.1,10.1-10.1C84.2,46.7,88.7,51.3,88.7,56.8z};
  }
}

\newcommand\orcidicon[1]{\href{https://orcid.org/#1}{\mbox{\scalerel*{
\begin{tikzpicture}[yscale=-1,transform shape]
\pic{orcidlogo};
\end{tikzpicture}
}{|}}}}

\usepackage{hyperref} %<--- Load after everything else

\definecolor{forestgreen}{rgb}{0.13, 0.55, 0.13}

\begin{document}

%%TC:ignore
\preprint{APS}

\title{The void halo mass function: a promising probe of neutrino mass}

\author{Gemma Zhang \orcidicon{0000-0002-8019-8082}}
\affiliation{Department of Physics, Princeton University, Princeton, NJ 08544, USA}
\affiliation{Department of Astrophysical Sciences, Princeton University, 4 Ivy Lane, Princeton, NJ 08544, USA}
\author{Zack Li \orcidicon{0000-0002-0309-9750}}
\affiliation{Department of Astrophysical Sciences, Princeton University, 4 Ivy Lane, Princeton, NJ 08544, USA}
\author{Jia Liu \orcidicon{0000-0001-8219-1995}}
\affiliation{Berkeley Center for Cosmological Physics, University of California, Berkeley, CA 94720, USA}
\affiliation{Lawrence Berkeley National Laboratory, 1 Cyclotron Road, Berkeley, CA 93720, USA}
\affiliation{Department of Astrophysical Sciences, Princeton University, 4 Ivy Lane, Princeton, NJ 08544, USA}
\author{David N. Spergel \orcidicon{0000-0002-5151-0006}}
\affiliation{Center for Computational Astrophysics, Flatiron Institute, 162 5th Avenue, 10010, New York, NY, USA}
\affiliation{Department of Astrophysical Sciences, Princeton University, 4 Ivy Lane, Princeton, NJ 08544, USA}
\author{Christina D. Kreisch \orcidicon{0000-0002-5061-7805}}
\affiliation{Department of Astrophysical Sciences, Princeton University, 4 Ivy Lane, Princeton, NJ 08544, USA}
\author{Alice Pisani \orcidicon{0000-0002-6146-4437}}
\affiliation{Department of Astrophysical Sciences, Princeton University, 4 Ivy Lane, Princeton, NJ 08544, USA}
\author{Benjamin D. Wandelt \orcidicon{0000-0002-5854-8269}}
\affiliation{Sorbonne Universit\'e, CNRS, UMR 7095, Institut d'Astrophysique de Paris, 98 bis boulevard Arago, 75014 Paris, France}
\affiliation{Sorbonne Universit\'es, Institut Lagrange de Paris, 98 bis boulevard Arago, 75014 Paris, France}
\affiliation{Center for Computational Astrophysics, Flatiron Institute, 162 5th Avenue, 10010, New York, NY, USA}

\date{\today}

\begin{abstract}
Cosmic voids, the underdense regions in the universe, are particularly sensitive to diffuse density components such as cosmic neutrinos. This sensitivity is enhanced by the match between void sizes and the free-streaming scale of massive neutrinos. Using the massive neutrino simulations \texttt{MassiveNuS}, we investigate the effect of neutrino mass on dark matter halos as a function of environment. We find that the halo mass function depends strongly on neutrino mass and that this dependence is more pronounced in voids than in high-density environments. An observational program that measured the characteristic mass of the most massive halos in voids should be able to place novel constraints on the sum of the masses of neutrinos $\sum m_\nu$. The neutrino mass effect in the simulations is quite strong: In  a 512$^3$  $h^{-3}$ Mpc$^3$ survey, the mean mass of the 1000 most massive halos in the void interiors is  $(4.82 \pm 0.11) \times 10^{12}~h^{-1}M_{\odot}$ for  $\sum m_\nu = 0.6$~eV and $(8.21 \pm 0.13) \times 10^{12}~h^{-1}M_{\odot}$ for $\sum m_\nu = 0.1$~eV. Subaru (SuMIRe), Euclid and WFIRST will have both spectroscopic and weak lensing surveys. Covering volumes at least 50 times larger than our simulations, they should be  sensitive probes of neutrino mass through void substructure.

%Cosmic voids, the underdense regions in the universe, are particularly sensitive to neutrinos. Using the \texttt{MassiveNuS} simulations, we find the halo mass function depends strongly on the sum of the neutrino masses $\sum m_{\nu}$, more so in voids than dense regions. In a 512$^3$  $h^{-3}$ Mpc$^3$ survey, the mean mass of the 1000 most massive halos in the voids is $(4.82 \pm 0.11) \times 10^{12}~h^{-1}M_{\odot}$ for $\sum m_\nu = 0.6$~eV and $(8.21 \pm 0.13) \times 10^{12}~h^{-1}M_{\odot}$ for $\sum m_\nu = 0.1$~eV. Future surveys like SuMIRe, Euclid and WFIRST will cover volumes at least 50 times larger than our simulations, strongly constraining $\sum m_{\nu}$ through void substructure.
\end{abstract}
                              
\keywords{cosmology$-$simulation; large-scale structure of the universe}
\maketitle

%%TC:endignore

\section{Introduction}

The masses of the neutrinos are the last unknown parameters in the Standard Model of particle physics. Particle physicists have measured the mass splitting between neutrino mass eigenstates~\citep{neutrinoosci, neutrinoosci2}; however, the total neutrino mass remains unknown. Cosmological observations can complement laboratory experiments by measuring the total mass of all three neutrino species, $\sum m_{\nu}$, through its effect on large-scale structures.  Within the context of the $\Lambda$CDM model, the current best constraint of $\sum m_{\nu}$ based on cosmological measurements is given by~\cite{planck} which concludes that $\sum m_{\nu} <$~0.12~$\rm{eV}$ (95\%~CL), a result derived from combined Planck Cosmic Microwave Background (CMB), CMB lensing, and baryon acoustic oscillations (BAO) information. More recently, simulation studies (such as \cite{massara2015, Kreisch2018,schuster2019}) demonstrate that cosmic voids are sensitive to massive neutrinos, which can be used as a new probe of the total neutrino mass. With the availability of an increasing amount of observational data from the sparse regions in the universe, voids offer new insights into neutrino properties complementary to other cosmological studies.

Voids have been used to probe various cosmological properties. Past works used void properties to constrain $\Omega_m$ and the linear growth rate of structures~\citep[see e.g.][]{hamaus2016, hawken2017, hamaus2017dwj, cai2016jek}. Studies of void properties are shedding light on our understanding of dark energy as the shape, number count, and density profile of voids have been shown to be sensitive to the properties of dark energy~\citep{park2007, lee2009, lavaux2010, biswas2010, sutter2012, stackedvoidsLW2012, bos2012, li2012, spolyar2013, sutter2015, pisani2015, sutter2015, pollina2016,  hamaus2016}.
In addition, because the environments inside voids are relatively simple compared to that of the densely clustered regions in the universe, void galaxies have a gentler history with less harassment, ram pressure stripping or other environmental effects~\citep{rojas2014, vandewaygaert2011, whitepaper,lee2009, hamaus2014jun, hamaus2014dec}. In this paper, we discuss how void substructure and, in particular, the dark matter halos in voids can potentially help constrain the total neutrino mass.  

Neutrinos are the only standard model particles that were relativistic during recombination but became non-relativistic during structure formation later in time~\citep{pdg}. The effect of neutrinos on structure growth depends on the free-streaming length, which is a function of their mass~\citep{lesgourgues2012}. At scales above their free-streaming length, neutrinos cluster the same way as cold dark matter and their masses have little effect on structure growth. However, at scales below their free-streaming length, they suppress the growth of structures. The free-streaming length of neutrinos is comparable to the size of voids, making voids attractive candidates for probing the neutrino mass observationally. For example, the neutrino free-streaming length is approximately 34~$h^{-1}$Mpc if $\sum m_{\nu}=0.6$~eV and 121~$h^{-1}$Mpc if $\sum m_{\nu}=0.1$~eV~\citep{lesgourgues2012}. 

In our study, we use N-body simulations with varying $\sum m_{\nu}$ to explore the effects of massive neutrinos on the halos within void. We separately search for halos in voids traced by dark matter halos and in voids traced by cold dark matter~(CDM) particles, which have been shown to exhibit different properties~\citep{Kreisch2018}. In order to verify our results, we perform the same analysis on the halo population of voids built from the halos and of voids built from the CDM particle field. In both cases, we find that the effect of massive neutrinos on the halo mass function in voids is more prominent than that on the population of all halos in the simulation. Thus, we show that halo mass information in voids has the potential to tighten the constraint on $\sum m_{\nu}$.   

In this paper, we begin by discussing the details of the N-body simulations, the void finder, and our method to search for void halos in Sec.~\ref{sec:analysis}. Then in Sec.~\ref{sec:results}, we present the results of the effects of massive neutrinos on halo mass functions of various halo populations. Finally, we discuss in Sec.~\ref{sec:conclusions} the implications of our results on future works and how they can be used in combination with observations to constrain the total neutrino mass.

\section{Analysis}
\label{sec:analysis}
In our work, we use halo information generated by the Cosmological Massive Neutrino Simulations \citep[\texttt{MassiveNuS},][]{jia2018} with different neutrino mass inputs. To locate the voids in \texttt{MassiveNuS}, we use the publicly available Void IDentification and Examination \citep[\texttt{VIDE},][]{vide} code.

The \texttt{MassiveNuS} consist of 101~N-body simulation cosmologies with varying values of $\sum m_{\nu}$, the matter density today $\Omega_m$, and the primordial power spectrum fluctuation amplitude $A_s$~\citep{jia2018}. The simulations assume a flat-$\Lambda \rm{CDM}$ universe. In our analysis, we use two fiducial simulations:~$\sum m_{\nu}=0.1~\rm{eV}$ and $\sum m_\nu = ~0.6~\rm{eV}$.  All other cosmological parameters are fixed at $A_s = 2.1 \times 10^{-9}$, $\Omega_b = 0.05$, $h = 0.7$, and $w = -1$. Each simulation contains $1024^3$~particles in a  (512~$h^{-1}$Mpc)$^3$ box  with periodic boundary conditions. The simulation code adopted a modified version of the tree-particle Mesh code \texttt{Gadget-2}~\citep{gadget2}, where neutrinos are treated as a linear perturbation term to the matter density, but its evolution is sourced by the full nonlinear matter perturbation. We use the snapshots and halo catalogues at $z = 0$ and $z = 1$ in our analysis. 

The publicly available void finder \texttt{VIDE} is used to identify the voids in each simulation. \texttt{VIDE} performs Voronoi tessellation and watershed transform on the CDM particle distribution or the halo distribution to find voids~\citep{vide}. \texttt{VIDE} makes no assumption of spherical shapes for voids, but for simplicity, we quote a radius for each void, which is the radius of a sphere with an equivalent volume. To reduce the computational cost associated with running the void finder on the full CDM particle distribution, we subsample 1\% of the particles when tracing voids with CDM particles. Similar to the parent halo and sub-halo relation, large voids may contain small voids. To avoid overlapping voids, we only include the top-level voids (``level''$=0$ in the \texttt{VIDE} catalogue). These parent voids contain sub-voids that have been merged with the watershed transform, whenever the ridge-lines separating two sub-voids have densities below 0.2 times the mean particle density.

Next, we search for halos within half of the radius from each void center, and hence exclude the halos near void walls that may experience complicated galaxy formation processes. 
We refer to these halos as ``void halos.'' We study the halo mass function of these void halos and that of the whole halo population (``all halos" hereafter), for the two massive neutrino cosmologies of $\sum m_{\nu} = 0.1, 0.6$~eV. We then compare the mean virial mass~($\langle M_{vir} \rangle$) of the $N$ most massive halos, where we vary $N$ from 100 to 1000. 

\section{Results}
\label{sec:results}
For both all halos and void halos, we find that increasing the neutrino mass leads to a decrease in the number of massive halos, which is expected as massive neutrinos suppress the growth of structures. More importantly, this effect is more pronounced in the void halo population than in all halos at late time. Thus, we propose the halos found in void interiors to be a promising tool to  constrain the neutrino mass sum. 

\begin{figure*} [htp!]
    \centering
    \includegraphics[width=0.9\linewidth]{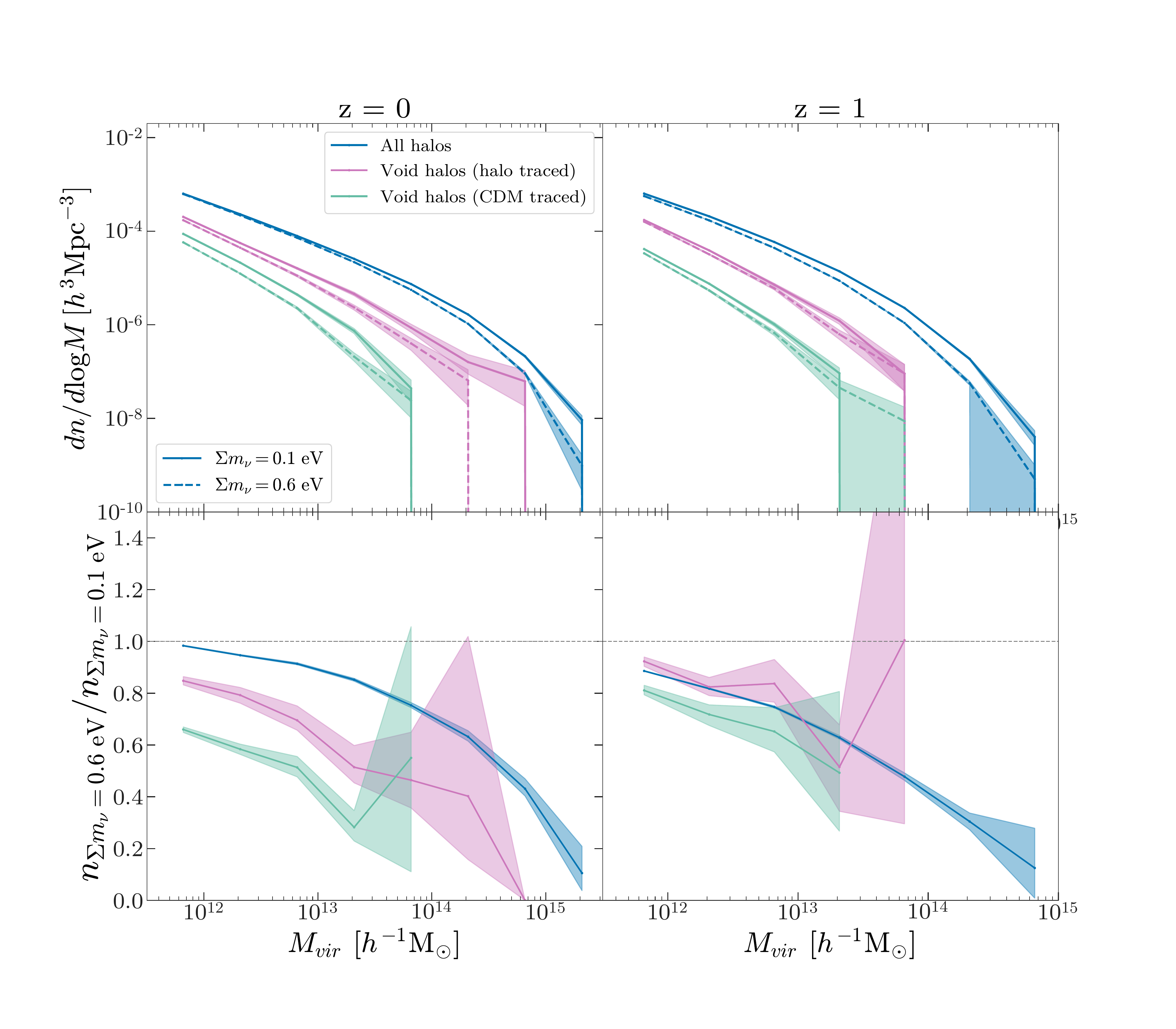}
    \caption{{\bf Upper panels}:  Halo mass functions of all halos (blue) and of void halos (halo-traced voids: magenta; CDM particle-traced voids: green) in the $\sum m_{\nu}=0.1$~eV (solid lines) and $\sum m_{\nu}=0.6$~eV (dashed line) simulations at $z=0$ (left) and $z=1$ (right). {\bf Lower panels}: Ratios of the halo mass functions from the $\sum m_{\nu}=0.6$~eV simulation to that of the $\sum m_{\nu}=0.1$~eV simulation for the three halo populations. The void halos consist of halos found within half of the void radii from the void centers.}
    \label{fig:trimmed_HMF_h}
\end{figure*}

In Fig.~\ref{fig:trimmed_HMF_h}, we show the halo mass function
~\citep{HMFpress,HMFbond} of all halos and of the void halos, for two different values of neutrino mass sum. The halo mass functions are normalized using (1)~the volume of the simulation box for all halos and (2)~the total volume of spheres of half void radii, as used in our void halo search, for the void halos. We show the Poisson error $\sqrt{\rm{N_{bin}}}$ where $\rm{N_{bin}}$ is the number count in each bin normalized by the respective volume. To examine the effect of massive neutrinos, in the lower panels of Fig.~\ref{fig:trimmed_HMF_h}, we plot the ratios of the halo mass function of $\sum m_{\nu}=0.6~\rm{eV}$ to that of $\sum m_{\nu}=0.1~\rm{eV}$, for both all halos and void halos at $z=0$ and $z=1$. The shaded regions represent the 68\% credible regions\footnote{To determine the errors in the ratio plots (lower panels), we randomly draw 200 samples from a Gaussian distribution with the same mean and standard deviation as the halo mass function, for each halo population, and then take the same ratios. The lower and upper errors are the 16 and 84 percentiles in the distribution, respectively}. 

\begin{figure*} [htp!]
    \centering
    \includegraphics[width=0.9\linewidth]{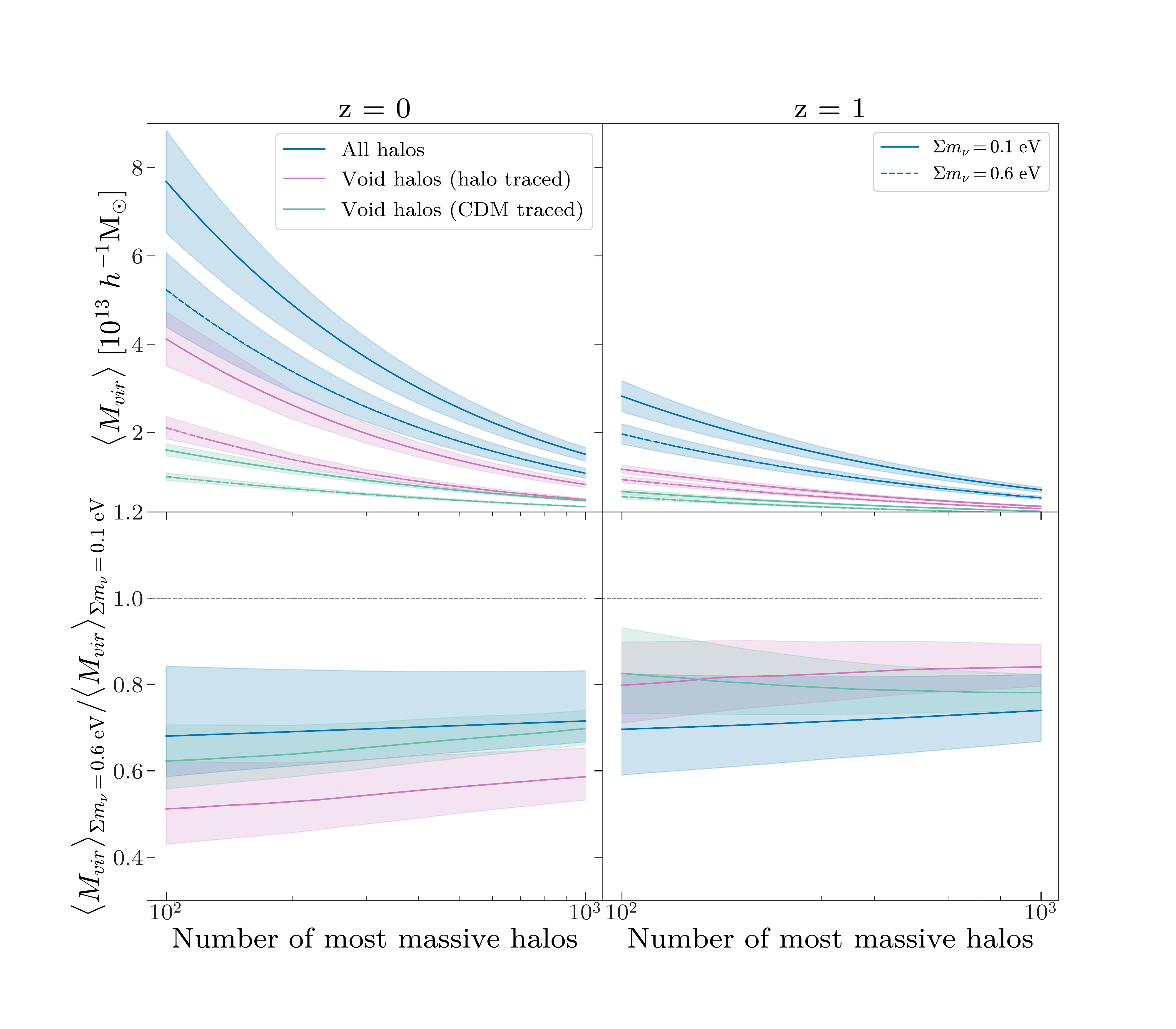}
    \caption{{\bf Upper panels}:  The mean mass of the $N$ most massive halos as a function of $N$ for all halos (blue) and void halos (halo-traced voids: magenta; CDM particle-traced voids: green) in the $\sum m_{\nu}=0.1$~eV (solid lines) and $\sum m_{\nu}=0.6$~eV (dashed line) simulations at $z=0$ (left) and $z=1$ (right). {\bf Lower panels}:  The ratio of the mean masses in the $\sum m_{\nu}=0.6$~eV simulation to that in the $\sum m_{\nu}=0.1$~eV simulation. The void halos consist of halos found within half of the void radii from the void centers.}
    \label{fig:trimmed_Nmassive_h}
\end{figure*}

While we see a decrease in the number density of halos as a function of mass for both all halos and void halos, this trend is more significant for void halos than for all halos at $z=0$. Neutrinos consist of a larger fraction of the matter density in voids than in denser regions, and this difference is more significant at $z=0$ than at $z=1$  as the neutrino mass effect is accumulated across time. We test our results using voids traced by halos and by CDM particles. The halo-traced voids are more accessible in observations, but are impacted by halo bias which can mimic the neutrino effect~\citep{pollina2016,Kreisch2018}. We find the same suppression effect on the halos in CDM particle-traced voids as in halo-traced voids, and the amplitude is stronger in the former population, verifying that our findings are not an artifact of halo bias.

Observationally, the halo mass function for void halos is difficult to determine due to their scarcity. Therefore, we propose a simpler statistic --- the average mass of $N$ most massive halos $\langle M_{vir} \rangle$. In practice, we would measure the average mass of the $N$ most luminous halos and rely on the relatively small scatter between stellar luminosity and mass implied by the $D_n$--$\sigma$ relation and the Tully-Fisher relation. In Fig.~\ref{fig:trimmed_Nmassive_h}, we show the mean virial mass of the $N$ most massive halos as a function of $N$, for all halos and void halos at $z=0$ and $z=1$. The shaded regions represent the 68\% credible regions\footnote{The errors for void halos are the standard deviation of 1,000 bootstrapped samples from the full void halo population. To obtain the error for all halos that are normalized to the same volume as the void halos, we first draw 100 random cubes in the simulation box, each with volume equivalent to the total void search volume. Within each cube, we then bootstrap 100 halo samples, each with the same number of halos as the total number of void halos. The errors are the standard deviation of the 10,000 resulting random samples. In the ratio plots (lower panels), The lower and upper errors are the 16 and 84 percentiles in the distributions, respectively.}. The reduction in the mean halo mass, due to the increase of neutrino mass from 0.1~eV to 0.6~eV, is larger in halo-traced void halos (e.g. by 50\% for $N=100$) than in all halos (by 30\% for $N=100$) at $z=0$. Our results demonstrate that void halos can potentially be a powerful tool for constraining the neutrino mass.

\section{Conclusions}
\label{sec:conclusions}
Using the halo and void information in \texttt{MassiveNuS}, we analyze the effect of neutrino mass on the mass function of void halos in comparison to that of the whole halo population. We performed the same analysis using two different void populations --- voids traced by cold dark matter particles and voids traced by halos. In both cases, our results show that the void halo mass function is more sensitive to neutrino mass than the halo mass function of the full simulation. The difference in the mass of void halos is shown to be highly statistically significant between $\sum m_{\nu}=0.1~\rm{eV}$ and $\sum m_{\nu}=0.6~\rm{eV}$ within our simulation box of volume $512^{3}~h^{-3}\rm{Mpc}^3$. For example, if we stack the 1000 most massive halos in void interiors from a survey of size $512^{3}~h^{-3}\rm{Mpc^3}$, the characteristic void halo mass should be $(4.82 \pm 0.11) \times 10^{12}~M_{\odot}$ for  $\sum m_\nu = 0.6$~eV and $(8.21 \pm 0.13) \times 10^{12}~M_{\odot}$ for $\sum m_\nu = 0.1$~eV. 

In summary, void halos present a promising way to probe the total mass of neutrinos. Observationally, voids can be identified using galaxy distributions, and the mass of halos within them can be measured from stacked weak lensing signals~\citep{lensing, stacked, stacked2}. Both will be available in high precision from upcoming surveys such as the Subaru Measurement of Images and Redshifts (SuMIRe), Wide Field Infrared Survey Telescope (WFIRST), and Euclid surveys. In this work, we explore the simple case of halos within half of the void radius. In the future, alternative definitions of void halos, such as halos within different radii or non-spherical boundaries, should be explored. To realize the potential of void halos, the degeneracy between neutrino mass and other physical effects such as baryons also needs to be carefully examined. In addition, we expect halos in voids to experience different and likely milder baryonic effects than those in overdense environments~\citep{Paillas2017}, and hence they can be a valuable consistency check against results obtained using the general halo population as typically done. 

\section*{Acknowledgements}
Part of this research was funded by the Princeton Astrophysics Undergraduate Summer Research Program~(to GZ). For using \texttt{MassiveNuS} in our work, we thank the Columbia Lensing group\footnote{\url{http://columbialensing.org}} for making their suite of simulations available, and NSF for supporting the creation of those simulations through grant AST-1602663 (to JL) and XSEDE allocation AST-140041; we also thank New Mexico State University (USA) and Instituto de Astrofisica de Andalucia CSIC (Spain) for hosting the Skies \& Universes site for cosmological simulation products. The Center for Computational Astrophysics is supported by the Simons Foundation.
CDK is supported by the National Science Foundation Graduate Research Fellowship under Grant DGE 1656466. AP is supported by NASA grant 15-WFIRST15-0008 to the WFIRST Science Investigation Team ``Cosmology with the High Latitude Survey''.
This work uses voids identified with \texttt{VIDE}\footnote{\url{http://www.cosmicvoids.net}}, which implements an enhanced version of \texttt{ZOBOV} to construct voids with a watershed algorithm.

\bibliography{references} 

\begin{thebibliography}{38}%
\makeatletter
\providecommand \@ifxundefined [1]{%
 \@ifx{#1\undefined}
}%
\providecommand \@ifnum [1]{%
 \ifnum #1\expandafter \@firstoftwo
 \else \expandafter \@secondoftwo
 \fi
}%
\providecommand \@ifx [1]{%
 \ifx #1\expandafter \@firstoftwo
 \else \expandafter \@secondoftwo
 \fi
}%
\providecommand \natexlab [1]{#1}%
\providecommand \enquote  [1]{``#1''}%
\providecommand \bibnamefont  [1]{#1}%
\providecommand \bibfnamefont [1]{#1}%
\providecommand \citenamefont [1]{#1}%
\providecommand \href@noop [0]{\@secondoftwo}%
\providecommand \href [0]{\begingroup \@sanitize@url \@href}%
\providecommand \@href[1]{\@@startlink{#1}\@@href}%
\providecommand \@@href[1]{\endgroup#1\@@endlink}%
\providecommand \@sanitize@url [0]{\catcode `\\12\catcode `\$12\catcode
  `\&12\catcode `\#12\catcode `\^12\catcode `\_12\catcode `\%12\relax}%
\providecommand \@@startlink[1]{}%
\providecommand \@@endlink[0]{}%
\providecommand \url  [0]{\begingroup\@sanitize@url \@url }%
\providecommand \@url [1]{\endgroup\@href {#1}{\urlprefix }}%
\providecommand \urlprefix  [0]{URL }%
\providecommand \Eprint [0]{\href }%
\providecommand \doibase [0]{https://doi.org/}%
\providecommand \selectlanguage [0]{\@gobble}%
\providecommand \bibinfo  [0]{\@secondoftwo}%
\providecommand \bibfield  [0]{\@secondoftwo}%
\providecommand \translation [1]{[#1]}%
\providecommand \BibitemOpen [0]{}%
\providecommand \bibitemStop [0]{}%
\providecommand \bibitemNoStop [0]{.\EOS\space}%
\providecommand \EOS [0]{\spacefactor3000\relax}%
\providecommand \BibitemShut  [1]{\csname bibitem#1\endcsname}%
\let\auto@bib@innerbib\@empty
%</preamble>
\bibitem [{\citenamefont {Fukuda}\ \emph {et~al.}(1998)\citenamefont {Fukuda},
  \citenamefont {Hayakawa}, \citenamefont {Ichihara}, \citenamefont {Inoue},
  \citenamefont {Ishihara}, \citenamefont {Ishino}, \citenamefont {Itow},
  \citenamefont {Kajita}, \citenamefont {Kameda}, \citenamefont {Kasuga},
  \citenamefont {Kobayashi}, \citenamefont {Kobayashi}, \citenamefont {Koshio},
  \citenamefont {Miura}, \citenamefont {Nakahata}, \citenamefont {Nakayama},
  \citenamefont {Okada}, \citenamefont {Okumura}, \citenamefont {Sakurai},
  \citenamefont {Shiozawa}, \citenamefont {Suzuki}, \citenamefont {Takeuchi},
  \citenamefont {Totsuka}, \citenamefont {Yamada}, \citenamefont {Earl},
  \citenamefont {Habig}, \citenamefont {Kearns}, \citenamefont {Messier},
  \citenamefont {Scholberg}, \citenamefont {Stone}, \citenamefont {Sulak},
  \citenamefont {Walter}, \citenamefont {Goldhaber}, \citenamefont
  {Barszczxak}, \citenamefont {Casper}, \citenamefont {Gajewski}, \citenamefont
  {Halverson}, \citenamefont {Hsu}, \citenamefont {Kropp}, \citenamefont
  {Price},\ and\ \citenamefont {Reines}}]{neutrinoosci}%
  \BibitemOpen
  \bibfield  {author} {\bibinfo {author} {\bibfnamefont {Y.}~\bibnamefont
  {Fukuda}}, \bibinfo {author} {\bibfnamefont {T.}~\bibnamefont {Hayakawa}},
  \bibinfo {author} {\bibfnamefont {E.}~\bibnamefont {Ichihara}}, \bibinfo
  {author} {\bibfnamefont {K.}~\bibnamefont {Inoue}}, \bibinfo {author}
  {\bibfnamefont {K.}~\bibnamefont {Ishihara}}, \bibinfo {author}
  {\bibfnamefont {H.}~\bibnamefont {Ishino}}, \bibinfo {author} {\bibfnamefont
  {Y.}~\bibnamefont {Itow}}, \bibinfo {author} {\bibfnamefont {T.}~\bibnamefont
  {Kajita}}, \bibinfo {author} {\bibfnamefont {J.}~\bibnamefont {Kameda}},
  \bibinfo {author} {\bibfnamefont {S.}~\bibnamefont {Kasuga}}, \bibinfo
  {author} {\bibfnamefont {K.}~\bibnamefont {Kobayashi}}, \bibinfo {author}
  {\bibfnamefont {Y.}~\bibnamefont {Kobayashi}}, \bibinfo {author}
  {\bibfnamefont {Y.}~\bibnamefont {Koshio}}, \bibinfo {author} {\bibfnamefont
  {M.}~\bibnamefont {Miura}}, \bibinfo {author} {\bibfnamefont
  {M.}~\bibnamefont {Nakahata}}, \bibinfo {author} {\bibfnamefont
  {S.}~\bibnamefont {Nakayama}}, \bibinfo {author} {\bibfnamefont
  {A.}~\bibnamefont {Okada}}, \bibinfo {author} {\bibfnamefont
  {K.}~\bibnamefont {Okumura}}, \bibinfo {author} {\bibfnamefont
  {N.}~\bibnamefont {Sakurai}}, \bibinfo {author} {\bibfnamefont
  {M.}~\bibnamefont {Shiozawa}}, \bibinfo {author} {\bibfnamefont
  {Y.}~\bibnamefont {Suzuki}}, \bibinfo {author} {\bibfnamefont
  {Y.}~\bibnamefont {Takeuchi}}, \bibinfo {author} {\bibfnamefont
  {Y.}~\bibnamefont {Totsuka}}, \bibinfo {author} {\bibfnamefont
  {S.}~\bibnamefont {Yamada}}, \bibinfo {author} {\bibfnamefont
  {M.}~\bibnamefont {Earl}}, \bibinfo {author} {\bibfnamefont {A.}~\bibnamefont
  {Habig}}, \bibinfo {author} {\bibfnamefont {E.}~\bibnamefont {Kearns}},
  \bibinfo {author} {\bibfnamefont {M.~D.}\ \bibnamefont {Messier}}, \bibinfo
  {author} {\bibfnamefont {K.}~\bibnamefont {Scholberg}}, \bibinfo {author}
  {\bibfnamefont {J.~L.}\ \bibnamefont {Stone}}, \bibinfo {author}
  {\bibfnamefont {L.~R.}\ \bibnamefont {Sulak}}, \bibinfo {author}
  {\bibfnamefont {C.~W.}\ \bibnamefont {Walter}}, \bibinfo {author}
  {\bibfnamefont {M.}~\bibnamefont {Goldhaber}}, \bibinfo {author}
  {\bibfnamefont {T.}~\bibnamefont {Barszczxak}}, \bibinfo {author}
  {\bibfnamefont {D.}~\bibnamefont {Casper}}, \bibinfo {author} {\bibfnamefont
  {W.}~\bibnamefont {Gajewski}}, \bibinfo {author} {\bibfnamefont {P.~G.}\
  \bibnamefont {Halverson}}, \bibinfo {author} {\bibfnamefont {J.}~\bibnamefont
  {Hsu}}, \bibinfo {author} {\bibfnamefont {W.~R.}\ \bibnamefont {Kropp}},
  \bibinfo {author} {\bibfnamefont {L.~R.}\ \bibnamefont {Price}},\ and\
  \bibinfo {author} {\bibfnamefont {F.}~\bibnamefont {Reines}} (\bibinfo
  {collaboration} {Super-Kamiokande Collaboration}),\ }\bibfield  {title}
  {\bibinfo {title} {Evidence for oscillation of atmospheric neutrinos},\
  }\href {https://doi.org/10.1103/PhysRevLett.81.1562} {\bibfield  {journal}
  {\bibinfo  {journal} {Phys. Rev. Lett.}\ }\textbf {\bibinfo {volume} {81}},\
  \bibinfo {pages} {1562} (\bibinfo {year} {1998})}\BibitemShut {NoStop}%
\bibitem [{\citenamefont {Ahmad}\ \emph {et~al.}(2002)\citenamefont {Ahmad},
  \citenamefont {Allen}, \citenamefont {Andersen}, \citenamefont {D.Anglin},
  \citenamefont {Barton}, \citenamefont {Beier}, \citenamefont {Bercovitch},
  \citenamefont {Bigu}, \citenamefont {Biller}, \citenamefont {Black},
  \citenamefont {Blevis}, \citenamefont {Boardman}, \citenamefont {Boger},
  \citenamefont {Bonvin}, \citenamefont {Boulay}, \citenamefont {Bowler},
  \citenamefont {Bowles}, \citenamefont {Brice}, \citenamefont {Browne},
  \citenamefont {Bullard}, \citenamefont {B\"uhler},\ and\ \citenamefont
  {Cameron}}]{neutrinoosci2}%
  \BibitemOpen
  \bibfield  {author} {\bibinfo {author} {\bibfnamefont {Q.~R.}\ \bibnamefont
  {Ahmad}}, \bibinfo {author} {\bibfnamefont {R.~C.}\ \bibnamefont {Allen}},
  \bibinfo {author} {\bibfnamefont {T.~C.}\ \bibnamefont {Andersen}}, \bibinfo
  {author} {\bibfnamefont {J.}~\bibnamefont {D.Anglin}}, \bibinfo {author}
  {\bibfnamefont {J.~C.}\ \bibnamefont {Barton}}, \bibinfo {author}
  {\bibfnamefont {E.~W.}\ \bibnamefont {Beier}}, \bibinfo {author}
  {\bibfnamefont {M.}~\bibnamefont {Bercovitch}}, \bibinfo {author}
  {\bibfnamefont {J.}~\bibnamefont {Bigu}}, \bibinfo {author} {\bibfnamefont
  {S.~D.}\ \bibnamefont {Biller}}, \bibinfo {author} {\bibfnamefont {R.~A.}\
  \bibnamefont {Black}}, \bibinfo {author} {\bibfnamefont {I.}~\bibnamefont
  {Blevis}}, \bibinfo {author} {\bibfnamefont {R.~J.}\ \bibnamefont
  {Boardman}}, \bibinfo {author} {\bibfnamefont {J.}~\bibnamefont {Boger}},
  \bibinfo {author} {\bibfnamefont {E.}~\bibnamefont {Bonvin}}, \bibinfo
  {author} {\bibfnamefont {M.~G.}\ \bibnamefont {Boulay}}, \bibinfo {author}
  {\bibfnamefont {M.~G.}\ \bibnamefont {Bowler}}, \bibinfo {author}
  {\bibfnamefont {T.~J.}\ \bibnamefont {Bowles}}, \bibinfo {author}
  {\bibfnamefont {S.~J.}\ \bibnamefont {Brice}}, \bibinfo {author}
  {\bibfnamefont {M.~C.}\ \bibnamefont {Browne}}, \bibinfo {author}
  {\bibfnamefont {T.~V.}\ \bibnamefont {Bullard}}, \bibinfo {author}
  {\bibfnamefont {G.}~\bibnamefont {B\"uhler}},\ and\ \bibinfo {author}
  {\bibfnamefont {J.}~\bibnamefont {Cameron}} (\bibinfo {collaboration} {SNO
  Collaboration}),\ }\bibfield  {title} {\bibinfo {title} {Direct evidence for
  neutrino flavor transformation from neutral-current interactions in the
  sudbury neutrino observatory},\ }\href
  {https://doi.org/10.1103/PhysRevLett.89.011301} {\bibfield  {journal}
  {\bibinfo  {journal} {Phys. Rev. Lett.}\ }\textbf {\bibinfo {volume} {89}},\
  \bibinfo {pages} {011301} (\bibinfo {year} {2002})}\BibitemShut {NoStop}%
\bibitem [{\citenamefont {{Planck Collaboration}}\ \emph
  {et~al.}(2018)\citenamefont {{Planck Collaboration}}, \citenamefont
  {{Aghanim}}, \citenamefont {{Akrami}}, \citenamefont {{Ashdown}},
  \citenamefont {{Aumont}}, \citenamefont {{Baccigalupi}}, \citenamefont
  {{Ballardini}}, \citenamefont {{Banday}}, \citenamefont {{Barreiro}},
  \citenamefont {{Bartolo}}, \citenamefont {{Basak}}, \citenamefont {{Battye}},
  \citenamefont {{Benabed}},\ and\ \citenamefont {{Bernard}}}]{planck}%
  \BibitemOpen
  \bibfield  {author} {\bibinfo {author} {\bibnamefont {{Planck
  Collaboration}}}, \bibinfo {author} {\bibfnamefont {N.}~\bibnamefont
  {{Aghanim}}}, \bibinfo {author} {\bibfnamefont {Y.}~\bibnamefont {{Akrami}}},
  \bibinfo {author} {\bibfnamefont {M.}~\bibnamefont {{Ashdown}}}, \bibinfo
  {author} {\bibfnamefont {J.}~\bibnamefont {{Aumont}}}, \bibinfo {author}
  {\bibfnamefont {C.}~\bibnamefont {{Baccigalupi}}}, \bibinfo {author}
  {\bibfnamefont {M.}~\bibnamefont {{Ballardini}}}, \bibinfo {author}
  {\bibfnamefont {A.~J.}\ \bibnamefont {{Banday}}}, \bibinfo {author}
  {\bibfnamefont {R.~B.}\ \bibnamefont {{Barreiro}}}, \bibinfo {author}
  {\bibfnamefont {N.}~\bibnamefont {{Bartolo}}}, \bibinfo {author}
  {\bibfnamefont {S.}~\bibnamefont {{Basak}}}, \bibinfo {author} {\bibfnamefont
  {R.}~\bibnamefont {{Battye}}}, \bibinfo {author} {\bibfnamefont
  {K.}~\bibnamefont {{Benabed}}},\ and\ \bibinfo {author} {\bibfnamefont
  {J.-P.}\ \bibnamefont {{Bernard}}},\ }\bibfield  {title} {\bibinfo {title}
  {{Planck 2018 results. VI. Cosmological parameters}},\ }\href@noop {} {\
  (\bibinfo {year} {2018})},\ \Eprint {https://arxiv.org/abs/1807.06209}
  {arXiv:1807.06209 [astro-ph.CO]} \BibitemShut {NoStop}%
%%CITATION = ARXIV:1807.06209;%%
\bibitem [{\citenamefont {{Massara}}\ \emph {et~al.}(2015)\citenamefont
  {{Massara}}, \citenamefont {{Villaescusa-Navarro}}, \citenamefont {{Viel}},\
  and\ \citenamefont {{Sutter}}}]{massara2015}%
  \BibitemOpen
  \bibfield  {author} {\bibinfo {author} {\bibfnamefont {E.}~\bibnamefont
  {{Massara}}}, \bibinfo {author} {\bibfnamefont {F.}~\bibnamefont
  {{Villaescusa-Navarro}}}, \bibinfo {author} {\bibfnamefont {M.}~\bibnamefont
  {{Viel}}},\ and\ \bibinfo {author} {\bibfnamefont {P.~M.}\ \bibnamefont
  {{Sutter}}},\ }\bibfield  {title} {\bibinfo {title} {{Voids in massive
  neutrino cosmologies}},\ }\href
  {https://doi.org/10.1088/1475-7516/2015/11/018} {\bibfield  {journal}
  {\bibinfo  {journal} {\jcap}\ }\textbf {\bibinfo {volume} {2015}},\ \bibinfo
  {eid} {018} (\bibinfo {year} {2015})},\ \Eprint
  {https://arxiv.org/abs/1506.03088} {arXiv:1506.03088 [astro-ph.CO]}
  \BibitemShut {NoStop}%
\bibitem [{\citenamefont {{Kreisch}}\ \emph {et~al.}(2019)\citenamefont
  {{Kreisch}}, \citenamefont {{Pisani}}, \citenamefont {{Carbone}},
  \citenamefont {{Liu}}, \citenamefont {{Hawken}}, \citenamefont {{Massara}},
  \citenamefont {{Spergel}},\ and\ \citenamefont {{Wandelt}}}]{Kreisch2018}%
  \BibitemOpen
  \bibfield  {author} {\bibinfo {author} {\bibfnamefont {C.~D.}\ \bibnamefont
  {{Kreisch}}}, \bibinfo {author} {\bibfnamefont {A.}~\bibnamefont {{Pisani}}},
  \bibinfo {author} {\bibfnamefont {C.}~\bibnamefont {{Carbone}}}, \bibinfo
  {author} {\bibfnamefont {J.}~\bibnamefont {{Liu}}}, \bibinfo {author}
  {\bibfnamefont {A.~J.}\ \bibnamefont {{Hawken}}}, \bibinfo {author}
  {\bibfnamefont {E.}~\bibnamefont {{Massara}}}, \bibinfo {author}
  {\bibfnamefont {D.~N.}\ \bibnamefont {{Spergel}}},\ and\ \bibinfo {author}
  {\bibfnamefont {B.~D.}\ \bibnamefont {{Wandelt}}},\ }\bibfield  {title}
  {\bibinfo {title} {{Massive neutrinos leave fingerprints on cosmic voids}},\
  }\href {https://doi.org/10.1093/mnras/stz1944} {\bibfield  {journal}
  {\bibinfo  {journal} {\mnras}\ }\textbf {\bibinfo {volume} {488}},\ \bibinfo
  {pages} {4413} (\bibinfo {year} {2019})},\ \Eprint
  {https://arxiv.org/abs/1808.07464} {arXiv:1808.07464 [astro-ph.CO]}
  \BibitemShut {NoStop}%
\bibitem [{\citenamefont {{Schuster}}\ \emph {et~al.}(2019)\citenamefont
  {{Schuster}}, \citenamefont {{Hamaus}}, \citenamefont {{Pisani}},
  \citenamefont {{Carbone}}, \citenamefont {{Kreisch}}, \citenamefont
  {{Pollina}},\ and\ \citenamefont {{Weller}}}]{schuster2019}%
  \BibitemOpen
  \bibfield  {author} {\bibinfo {author} {\bibfnamefont {N.}~\bibnamefont
  {{Schuster}}}, \bibinfo {author} {\bibfnamefont {N.}~\bibnamefont
  {{Hamaus}}}, \bibinfo {author} {\bibfnamefont {A.}~\bibnamefont {{Pisani}}},
  \bibinfo {author} {\bibfnamefont {C.}~\bibnamefont {{Carbone}}}, \bibinfo
  {author} {\bibfnamefont {C.~D.}\ \bibnamefont {{Kreisch}}}, \bibinfo {author}
  {\bibfnamefont {G.}~\bibnamefont {{Pollina}}},\ and\ \bibinfo {author}
  {\bibfnamefont {J.}~\bibnamefont {{Weller}}},\ }\bibfield  {title} {\bibinfo
  {title} {{The bias of cosmic voids in the presence of massive neutrinos}},\
  }\href@noop {} {\bibfield  {journal} {\bibinfo  {journal} {arXiv e-prints}\
  ,\ \bibinfo {eid} {arXiv:1905.00436}} (\bibinfo {year} {2019})},\ \Eprint
  {https://arxiv.org/abs/1905.00436} {arXiv:1905.00436 [astro-ph.CO]}
  \BibitemShut {NoStop}%
\bibitem [{\citenamefont {Hamaus}\ \emph {et~al.}(2016)\citenamefont {Hamaus},
  \citenamefont {Pisani}, \citenamefont {Sutter}, \citenamefont {Lavaux},
  \citenamefont {Escoffier}, \citenamefont {Wandelt},\ and\ \citenamefont
  {Weller}}]{hamaus2016}%
  \BibitemOpen
  \bibfield  {author} {\bibinfo {author} {\bibfnamefont {N.}~\bibnamefont
  {Hamaus}}, \bibinfo {author} {\bibfnamefont {A.}~\bibnamefont {Pisani}},
  \bibinfo {author} {\bibfnamefont {P.~M.}\ \bibnamefont {Sutter}}, \bibinfo
  {author} {\bibfnamefont {G.}~\bibnamefont {Lavaux}}, \bibinfo {author}
  {\bibfnamefont {S.}~\bibnamefont {Escoffier}}, \bibinfo {author}
  {\bibfnamefont {B.~D.}\ \bibnamefont {Wandelt}},\ and\ \bibinfo {author}
  {\bibfnamefont {J.}~\bibnamefont {Weller}},\ }\bibfield  {title} {\bibinfo
  {title} {Constraints on cosmology and gravity from the dynamics of voids},\
  }\href {https://doi.org/10.1103/PhysRevLett.117.091302} {\bibfield  {journal}
  {\bibinfo  {journal} {Phys. Rev. Lett.}\ }\textbf {\bibinfo {volume} {117}},\
  \bibinfo {pages} {091302} (\bibinfo {year} {2016})}\BibitemShut {NoStop}%
\bibitem [{\citenamefont {{Hawken}}\ \emph {et~al.}(2017)\citenamefont
  {{Hawken}}, \citenamefont {{Granett}}, \citenamefont {{Iovino}},
  \citenamefont {{Guzzo}}, \citenamefont {{Peacock}}, \citenamefont {{de la
  Torre}}, \citenamefont {{Garilli}}, \citenamefont {{Bolzonella}},
  \citenamefont {{Scodeggio}}, \citenamefont {{Abbas}}, \citenamefont
  {{Adami}}, \citenamefont {{Bottini}}, \citenamefont {{Cappi}}, \citenamefont
  {{Cucciati}}, \citenamefont {{Davidzon}}, \citenamefont {{Fritz}},
  \citenamefont {{Franzetti}}, \citenamefont {{Krywult}}, \citenamefont {{Le
  Brun}}, \citenamefont {{Le F{\`e}vre}}, \citenamefont {{Maccagni}},
  \citenamefont {{Ma{\l}ek}}, \citenamefont {{Marulli}}, \citenamefont
  {{Polletta}}, \citenamefont {{Pollo}}, \citenamefont {{Tasca}}, \citenamefont
  {{Tojeiro}}, \citenamefont {{Vergani}}, \citenamefont {{Zanichelli}},
  \citenamefont {{Arnouts}}, \citenamefont {{Bel}}, \citenamefont
  {{Branchini}}, \citenamefont {{De Lucia}}, \citenamefont {{Ilbert}},
  \citenamefont {{Moscardini}},\ and\ \citenamefont {{Percival}}}]{hawken2017}%
  \BibitemOpen
  \bibfield  {author} {\bibinfo {author} {\bibfnamefont {A.~J.}\ \bibnamefont
  {{Hawken}}}, \bibinfo {author} {\bibfnamefont {B.~R.}\ \bibnamefont
  {{Granett}}}, \bibinfo {author} {\bibfnamefont {A.}~\bibnamefont {{Iovino}}},
  \bibinfo {author} {\bibfnamefont {L.}~\bibnamefont {{Guzzo}}}, \bibinfo
  {author} {\bibfnamefont {J.~A.}\ \bibnamefont {{Peacock}}}, \bibinfo {author}
  {\bibfnamefont {S.}~\bibnamefont {{de la Torre}}}, \bibinfo {author}
  {\bibfnamefont {B.}~\bibnamefont {{Garilli}}}, \bibinfo {author}
  {\bibfnamefont {M.}~\bibnamefont {{Bolzonella}}}, \bibinfo {author}
  {\bibfnamefont {M.}~\bibnamefont {{Scodeggio}}}, \bibinfo {author}
  {\bibfnamefont {U.}~\bibnamefont {{Abbas}}}, \bibinfo {author} {\bibfnamefont
  {C.}~\bibnamefont {{Adami}}}, \bibinfo {author} {\bibfnamefont
  {D.}~\bibnamefont {{Bottini}}}, \bibinfo {author} {\bibfnamefont
  {A.}~\bibnamefont {{Cappi}}}, \bibinfo {author} {\bibfnamefont
  {O.}~\bibnamefont {{Cucciati}}}, \bibinfo {author} {\bibfnamefont
  {I.}~\bibnamefont {{Davidzon}}}, \bibinfo {author} {\bibfnamefont
  {A.}~\bibnamefont {{Fritz}}}, \bibinfo {author} {\bibfnamefont
  {P.}~\bibnamefont {{Franzetti}}}, \bibinfo {author} {\bibfnamefont
  {J.}~\bibnamefont {{Krywult}}}, \bibinfo {author} {\bibfnamefont
  {V.}~\bibnamefont {{Le Brun}}}, \bibinfo {author} {\bibfnamefont
  {O.}~\bibnamefont {{Le F{\`e}vre}}}, \bibinfo {author} {\bibfnamefont
  {D.}~\bibnamefont {{Maccagni}}}, \bibinfo {author} {\bibfnamefont
  {K.}~\bibnamefont {{Ma{\l}ek}}}, \bibinfo {author} {\bibfnamefont
  {F.}~\bibnamefont {{Marulli}}}, \bibinfo {author} {\bibfnamefont
  {M.}~\bibnamefont {{Polletta}}}, \bibinfo {author} {\bibfnamefont
  {A.}~\bibnamefont {{Pollo}}}, \bibinfo {author} {\bibfnamefont {L.~A.~M.}\
  \bibnamefont {{Tasca}}}, \bibinfo {author} {\bibfnamefont {R.}~\bibnamefont
  {{Tojeiro}}}, \bibinfo {author} {\bibfnamefont {D.}~\bibnamefont
  {{Vergani}}}, \bibinfo {author} {\bibfnamefont {A.}~\bibnamefont
  {{Zanichelli}}}, \bibinfo {author} {\bibfnamefont {S.}~\bibnamefont
  {{Arnouts}}}, \bibinfo {author} {\bibfnamefont {J.}~\bibnamefont {{Bel}}},
  \bibinfo {author} {\bibfnamefont {E.}~\bibnamefont {{Branchini}}}, \bibinfo
  {author} {\bibfnamefont {G.}~\bibnamefont {{De Lucia}}}, \bibinfo {author}
  {\bibfnamefont {O.}~\bibnamefont {{Ilbert}}}, \bibinfo {author}
  {\bibfnamefont {L.}~\bibnamefont {{Moscardini}}},\ and\ \bibinfo {author}
  {\bibfnamefont {W.~J.}\ \bibnamefont {{Percival}}},\ }\bibfield  {title}
  {\bibinfo {title} {{The VIMOS Public Extragalactic Redshift Survey. Measuring
  the growth rate of structure around cosmic voids}},\ }\href
  {https://doi.org/10.1051/0004-6361/201629678} {\bibfield  {journal} {\bibinfo
   {journal} {\aap}\ }\textbf {\bibinfo {volume} {607}},\ \bibinfo {eid} {A54}
  (\bibinfo {year} {2017})},\ \Eprint {https://arxiv.org/abs/1611.07046}
  {arXiv:1611.07046 [astro-ph.CO]} \BibitemShut {NoStop}%
\bibitem [{\citenamefont {Hamaus}\ \emph {et~al.}(2017)\citenamefont {Hamaus},
  \citenamefont {Cousinou}, \citenamefont {Pisani}, \citenamefont {Aubert},
  \citenamefont {Escoffier},\ and\ \citenamefont {Weller}}]{hamaus2017dwj}%
  \BibitemOpen
  \bibfield  {author} {\bibinfo {author} {\bibfnamefont {N.}~\bibnamefont
  {Hamaus}}, \bibinfo {author} {\bibfnamefont {M.-C.}\ \bibnamefont
  {Cousinou}}, \bibinfo {author} {\bibfnamefont {A.}~\bibnamefont {Pisani}},
  \bibinfo {author} {\bibfnamefont {M.}~\bibnamefont {Aubert}}, \bibinfo
  {author} {\bibfnamefont {S.}~\bibnamefont {Escoffier}},\ and\ \bibinfo
  {author} {\bibfnamefont {J.}~\bibnamefont {Weller}},\ }\bibfield  {title}
  {\bibinfo {title} {{Multipole analysis of redshift-space distortions around
  cosmic voids}},\ }\href {https://doi.org/10.1088/1475-7516/2017/07/014}
  {\bibfield  {journal} {\bibinfo  {journal} {JCAP}\ }\textbf {\bibinfo
  {volume} {1707}}\bibfield  {number} {\bibinfo  {number} { (07)},\ \bibinfo
  {pages} {014}},\ }\Eprint {https://arxiv.org/abs/1705.05328}
  {arXiv:1705.05328 [astro-ph.CO]} \BibitemShut {NoStop}%
%%CITATION = ARXIV:1705.05328;%%
\bibitem [{\citenamefont {Cai}\ \emph {et~al.}(2016)\citenamefont {Cai},
  \citenamefont {Taylor}, \citenamefont {Peacock},\ and\ \citenamefont
  {Padilla}}]{cai2016jek}%
  \BibitemOpen
  \bibfield  {author} {\bibinfo {author} {\bibfnamefont {Y.-C.}\ \bibnamefont
  {Cai}}, \bibinfo {author} {\bibfnamefont {A.}~\bibnamefont {Taylor}},
  \bibinfo {author} {\bibfnamefont {J.~A.}\ \bibnamefont {Peacock}},\ and\
  \bibinfo {author} {\bibfnamefont {N.}~\bibnamefont {Padilla}},\ }\bibfield
  {title} {\bibinfo {title} {{Redshift-space distortions around voids}},\
  }\href {https://doi.org/10.1093/mnras/stw1809} {\bibfield  {journal}
  {\bibinfo  {journal} {Mon. Not. Roy. Astron. Soc.}\ }\textbf {\bibinfo
  {volume} {462}},\ \bibinfo {pages} {2465} (\bibinfo {year} {2016})},\ \Eprint
  {https://arxiv.org/abs/1603.05184} {arXiv:1603.05184 [astro-ph.CO]}
  \BibitemShut {NoStop}%
%%CITATION = ARXIV:1603.05184;%%
\bibitem [{\citenamefont {{Park}}\ and\ \citenamefont
  {{Lee}}(2007)}]{park2007}%
  \BibitemOpen
  \bibfield  {author} {\bibinfo {author} {\bibfnamefont {D.}~\bibnamefont
  {{Park}}}\ and\ \bibinfo {author} {\bibfnamefont {J.}~\bibnamefont {{Lee}}},\
  }\bibfield  {title} {\bibinfo {title} {{Void Ellipticity Distribution as a
  Probe of Cosmology}},\ }\href {https://doi.org/10.1103/PhysRevLett.98.081301}
  {\bibfield  {journal} {\bibinfo  {journal} {\prl}\ }\textbf {\bibinfo
  {volume} {98}},\ \bibinfo {eid} {081301} (\bibinfo {year} {2007})},\ \Eprint
  {https://arxiv.org/abs/astro-ph/0610520} {arXiv:astro-ph/0610520 [astro-ph]}
  \BibitemShut {NoStop}%
\bibitem [{\citenamefont {{Lee}}\ and\ \citenamefont {{Park}}(2009)}]{lee2009}%
  \BibitemOpen
  \bibfield  {author} {\bibinfo {author} {\bibfnamefont {J.}~\bibnamefont
  {{Lee}}}\ and\ \bibinfo {author} {\bibfnamefont {D.}~\bibnamefont {{Park}}},\
  }\bibfield  {title} {\bibinfo {title} {{Constraining the Dark Energy Equation
  of State with Cosmic Voids}},\ }\href
  {https://doi.org/10.1088/0004-637X/696/1/L10} {\bibfield  {journal} {\bibinfo
   {journal} {\apjl}\ }\textbf {\bibinfo {volume} {696}},\ \bibinfo {pages}
  {L10} (\bibinfo {year} {2009})},\ \Eprint {https://arxiv.org/abs/0704.0881}
  {arXiv:0704.0881 [astro-ph]} \BibitemShut {NoStop}%
\bibitem [{\citenamefont {{Lavaux}}\ and\ \citenamefont
  {{Wandelt}}(2010)}]{lavaux2010}%
  \BibitemOpen
  \bibfield  {author} {\bibinfo {author} {\bibfnamefont {G.}~\bibnamefont
  {{Lavaux}}}\ and\ \bibinfo {author} {\bibfnamefont {B.~D.}\ \bibnamefont
  {{Wandelt}}},\ }\bibfield  {title} {\bibinfo {title} {{Precision cosmology
  with voids: definition, methods, dynamics}},\ }\href
  {https://doi.org/10.1111/j.1365-2966.2010.16197.x} {\bibfield  {journal}
  {\bibinfo  {journal} {\mnras}\ }\textbf {\bibinfo {volume} {403}},\ \bibinfo
  {pages} {1392} (\bibinfo {year} {2010})},\ \Eprint
  {https://arxiv.org/abs/0906.4101} {arXiv:0906.4101 [astro-ph.CO]}
  \BibitemShut {NoStop}%
\bibitem [{\citenamefont {{Biswas}}\ \emph {et~al.}(2010)\citenamefont
  {{Biswas}}, \citenamefont {{Alizadeh}},\ and\ \citenamefont
  {{Wandelt}}}]{biswas2010}%
  \BibitemOpen
  \bibfield  {author} {\bibinfo {author} {\bibfnamefont {R.}~\bibnamefont
  {{Biswas}}}, \bibinfo {author} {\bibfnamefont {E.}~\bibnamefont
  {{Alizadeh}}},\ and\ \bibinfo {author} {\bibfnamefont {B.~D.}\ \bibnamefont
  {{Wandelt}}},\ }\bibfield  {title} {\bibinfo {title} {{Voids as a precision
  probe of dark energy}},\ }\href {https://doi.org/10.1103/PhysRevD.82.023002}
  {\bibfield  {journal} {\bibinfo  {journal} {\prd}\ }\textbf {\bibinfo
  {volume} {82}},\ \bibinfo {eid} {023002} (\bibinfo {year} {2010})},\ \Eprint
  {https://arxiv.org/abs/1002.0014} {arXiv:1002.0014 [astro-ph.CO]}
  \BibitemShut {NoStop}%
\bibitem [{\citenamefont {{Sutter}}\ \emph {et~al.}(2012)\citenamefont
  {{Sutter}}, \citenamefont {{Lavaux}}, \citenamefont {{Wandelt}},\ and\
  \citenamefont {{Weinberg}}}]{sutter2012}%
  \BibitemOpen
  \bibfield  {author} {\bibinfo {author} {\bibfnamefont {P.~M.}\ \bibnamefont
  {{Sutter}}}, \bibinfo {author} {\bibfnamefont {G.}~\bibnamefont {{Lavaux}}},
  \bibinfo {author} {\bibfnamefont {B.~D.}\ \bibnamefont {{Wandelt}}},\ and\
  \bibinfo {author} {\bibfnamefont {D.~H.}\ \bibnamefont {{Weinberg}}},\
  }\bibfield  {title} {\bibinfo {title} {{A First Application of the
  Alcock-Paczynski Test to Stacked Cosmic Voids}},\ }\href
  {https://doi.org/10.1088/0004-637X/761/2/187} {\bibfield  {journal} {\bibinfo
   {journal} {\apj}\ }\textbf {\bibinfo {volume} {761}},\ \bibinfo {eid} {187}
  (\bibinfo {year} {2012})},\ \Eprint {https://arxiv.org/abs/1208.1058}
  {arXiv:1208.1058 [astro-ph.CO]} \BibitemShut {NoStop}%
\bibitem [{\citenamefont {{Lavaux}}\ and\ \citenamefont
  {{Wandelt}}(2012)}]{stackedvoidsLW2012}%
  \BibitemOpen
  \bibfield  {author} {\bibinfo {author} {\bibfnamefont {G.}~\bibnamefont
  {{Lavaux}}}\ and\ \bibinfo {author} {\bibfnamefont {B.~D.}\ \bibnamefont
  {{Wandelt}}},\ }\bibfield  {title} {\bibinfo {title} {{Precision Cosmography
  with Stacked Voids}},\ }\href {https://doi.org/10.1088/0004-637X/754/2/109}
  {\bibfield  {journal} {\bibinfo  {journal} {\apj}\ }\textbf {\bibinfo
  {volume} {754}},\ \bibinfo {eid} {109} (\bibinfo {year} {2012})},\ \Eprint
  {https://arxiv.org/abs/1110.0345} {arXiv:1110.0345 [astro-ph.CO]}
  \BibitemShut {NoStop}%
\bibitem [{\citenamefont {{Bos}}\ \emph {et~al.}(2012)\citenamefont {{Bos}},
  \citenamefont {{van de Weygaert}}, \citenamefont {{Dolag}},\ and\
  \citenamefont {{Pettorino}}}]{bos2012}%
  \BibitemOpen
  \bibfield  {author} {\bibinfo {author} {\bibfnamefont {E.~G.~P.}\
  \bibnamefont {{Bos}}}, \bibinfo {author} {\bibfnamefont {R.}~\bibnamefont
  {{van de Weygaert}}}, \bibinfo {author} {\bibfnamefont {K.}~\bibnamefont
  {{Dolag}}},\ and\ \bibinfo {author} {\bibfnamefont {V.}~\bibnamefont
  {{Pettorino}}},\ }\bibfield  {title} {\bibinfo {title} {{The darkness that
  shaped the void: dark energy and cosmic voids}},\ }\href
  {https://doi.org/10.1111/j.1365-2966.2012.21478.x} {\bibfield  {journal}
  {\bibinfo  {journal} {\mnras}\ }\textbf {\bibinfo {volume} {426}},\ \bibinfo
  {pages} {440} (\bibinfo {year} {2012})},\ \Eprint
  {https://arxiv.org/abs/1205.4238} {arXiv:1205.4238 [astro-ph.CO]}
  \BibitemShut {NoStop}%
\bibitem [{\citenamefont {{Li}}\ \emph {et~al.}(2012)\citenamefont {{Li}},
  \citenamefont {{Zhao}},\ and\ \citenamefont {{Koyama}}}]{li2012}%
  \BibitemOpen
  \bibfield  {author} {\bibinfo {author} {\bibfnamefont {B.}~\bibnamefont
  {{Li}}}, \bibinfo {author} {\bibfnamefont {G.-B.}\ \bibnamefont {{Zhao}}},\
  and\ \bibinfo {author} {\bibfnamefont {K.}~\bibnamefont {{Koyama}}},\
  }\bibfield  {title} {\bibinfo {title} {{Haloes and voids in f(R) gravity}},\
  }\href {https://doi.org/10.1111/j.1365-2966.2012.20573.x} {\bibfield
  {journal} {\bibinfo  {journal} {\mnras}\ }\textbf {\bibinfo {volume} {421}},\
  \bibinfo {pages} {3481} (\bibinfo {year} {2012})},\ \Eprint
  {https://arxiv.org/abs/1111.2602} {arXiv:1111.2602 [astro-ph.CO]}
  \BibitemShut {NoStop}%
\bibitem [{\citenamefont {{Spolyar}}\ \emph {et~al.}(2013)\citenamefont
  {{Spolyar}}, \citenamefont {{Sahl{\'e}n}},\ and\ \citenamefont
  {{Silk}}}]{spolyar2013}%
  \BibitemOpen
  \bibfield  {author} {\bibinfo {author} {\bibfnamefont {D.}~\bibnamefont
  {{Spolyar}}}, \bibinfo {author} {\bibfnamefont {M.}~\bibnamefont
  {{Sahl{\'e}n}}},\ and\ \bibinfo {author} {\bibfnamefont {J.}~\bibnamefont
  {{Silk}}},\ }\bibfield  {title} {\bibinfo {title} {{Topology and Dark Energy:
  Testing Gravity in Voids}},\ }\href
  {https://doi.org/10.1103/PhysRevLett.111.241103} {\bibfield  {journal}
  {\bibinfo  {journal} {\prl}\ }\textbf {\bibinfo {volume} {111}},\ \bibinfo
  {eid} {241103} (\bibinfo {year} {2013})},\ \Eprint
  {https://arxiv.org/abs/1304.5239} {arXiv:1304.5239 [astro-ph.CO]}
  \BibitemShut {NoStop}%
\bibitem [{\citenamefont {{Sutter}}\ \emph
  {et~al.}(2015{\natexlab{a}})\citenamefont {{Sutter}}, \citenamefont
  {{Carlesi}}, \citenamefont {{Wandelt}},\ and\ \citenamefont
  {{Knebe}}}]{sutter2015}%
  \BibitemOpen
  \bibfield  {author} {\bibinfo {author} {\bibfnamefont {P.~M.}\ \bibnamefont
  {{Sutter}}}, \bibinfo {author} {\bibfnamefont {E.}~\bibnamefont {{Carlesi}}},
  \bibinfo {author} {\bibfnamefont {B.~D.}\ \bibnamefont {{Wandelt}}},\ and\
  \bibinfo {author} {\bibfnamefont {A.}~\bibnamefont {{Knebe}}},\ }\bibfield
  {title} {\bibinfo {title} {{On the observability of coupled dark energy with
  cosmic voids.}},\ }\href {https://doi.org/10.1093/mnrasl/slu155} {\bibfield
  {journal} {\bibinfo  {journal} {\mnras}\ }\textbf {\bibinfo {volume} {446}},\
  \bibinfo {pages} {L1} (\bibinfo {year} {2015}{\natexlab{a}})},\ \Eprint
  {https://arxiv.org/abs/1406.0511} {arXiv:1406.0511 [astro-ph.CO]}
  \BibitemShut {NoStop}%
\bibitem [{\citenamefont {Pisani}\ \emph {et~al.}(2015)\citenamefont {Pisani},
  \citenamefont {Sutter}, \citenamefont {Hamaus}, \citenamefont {Alizadeh},
  \citenamefont {Biswas}, \citenamefont {Wandelt},\ and\ \citenamefont
  {Hirata}}]{pisani2015}%
  \BibitemOpen
  \bibfield  {author} {\bibinfo {author} {\bibfnamefont {A.}~\bibnamefont
  {Pisani}}, \bibinfo {author} {\bibfnamefont {P.~M.}\ \bibnamefont {Sutter}},
  \bibinfo {author} {\bibfnamefont {N.}~\bibnamefont {Hamaus}}, \bibinfo
  {author} {\bibfnamefont {E.}~\bibnamefont {Alizadeh}}, \bibinfo {author}
  {\bibfnamefont {R.}~\bibnamefont {Biswas}}, \bibinfo {author} {\bibfnamefont
  {B.~D.}\ \bibnamefont {Wandelt}},\ and\ \bibinfo {author} {\bibfnamefont
  {C.~M.}\ \bibnamefont {Hirata}},\ }\bibfield  {title} {\bibinfo {title}
  {Counting voids to probe dark energy},\ }\href
  {https://doi.org/10.1103/PhysRevD.92.083531} {\bibfield  {journal} {\bibinfo
  {journal} {Phys. Rev. D}\ }\textbf {\bibinfo {volume} {92}},\ \bibinfo
  {pages} {083531} (\bibinfo {year} {2015})}\BibitemShut {NoStop}%
\bibitem [{\citenamefont {{Pollina}}\ \emph {et~al.}(2016)\citenamefont
  {{Pollina}}, \citenamefont {{Baldi}}, \citenamefont {{Marulli}},\ and\
  \citenamefont {{Moscardini}}}]{pollina2016}%
  \BibitemOpen
  \bibfield  {author} {\bibinfo {author} {\bibfnamefont {G.}~\bibnamefont
  {{Pollina}}}, \bibinfo {author} {\bibfnamefont {M.}~\bibnamefont {{Baldi}}},
  \bibinfo {author} {\bibfnamefont {F.}~\bibnamefont {{Marulli}}},\ and\
  \bibinfo {author} {\bibfnamefont {L.}~\bibnamefont {{Moscardini}}},\
  }\bibfield  {title} {\bibinfo {title} {{Cosmic voids in coupled dark energy
  cosmologies: the impact of halo bias}},\ }\href
  {https://doi.org/10.1093/mnras/stv2503} {\bibfield  {journal} {\bibinfo
  {journal} {\mnras}\ }\textbf {\bibinfo {volume} {455}},\ \bibinfo {pages}
  {3075} (\bibinfo {year} {2016})},\ \Eprint {https://arxiv.org/abs/1506.08831}
  {arXiv:1506.08831 [astro-ph.CO]} \BibitemShut {NoStop}%
\bibitem [{\citenamefont {{Rojas}}\ \emph {et~al.}(2004)\citenamefont
  {{Rojas}}, \citenamefont {{Vogeley}}, \citenamefont {{Hoyle}},\ and\
  \citenamefont {{Brinkmann}}}]{rojas2014}%
  \BibitemOpen
  \bibfield  {author} {\bibinfo {author} {\bibfnamefont {R.~R.}\ \bibnamefont
  {{Rojas}}}, \bibinfo {author} {\bibfnamefont {M.~S.}\ \bibnamefont
  {{Vogeley}}}, \bibinfo {author} {\bibfnamefont {F.}~\bibnamefont {{Hoyle}}},\
  and\ \bibinfo {author} {\bibfnamefont {J.}~\bibnamefont {{Brinkmann}}},\
  }\bibfield  {title} {\bibinfo {title} {{Photometric Properties of Void
  Galaxies in the Sloan Digital Sky Survey}},\ }\href
  {https://doi.org/10.1086/425225} {\bibfield  {journal} {\bibinfo  {journal}
  {\apj}\ }\textbf {\bibinfo {volume} {617}},\ \bibinfo {pages} {50} (\bibinfo
  {year} {2004})},\ \Eprint {https://arxiv.org/abs/astro-ph/0307274}
  {arXiv:astro-ph/0307274 [astro-ph]} \BibitemShut {NoStop}%
\bibitem [{\citenamefont {{van de Weygaert}}\ \emph {et~al.}(2011)\citenamefont
  {{van de Weygaert}}, \citenamefont {{Kreckel}}, \citenamefont {{Platen}},
  \citenamefont {{Beygu}}, \citenamefont {{van Gorkom}}, \citenamefont {{van
  der Hulst}}, \citenamefont {{Arag{\'o}n-Calvo}}, \citenamefont {{Peebles}},
  \citenamefont {{Jarrett}}, \citenamefont {{Rhee}}, \citenamefont
  {{Kova{\v{c}}}},\ and\ \citenamefont {{Yip}}}]{vandewaygaert2011}%
  \BibitemOpen
  \bibfield  {author} {\bibinfo {author} {\bibfnamefont {R.}~\bibnamefont {{van
  de Weygaert}}}, \bibinfo {author} {\bibfnamefont {K.}~\bibnamefont
  {{Kreckel}}}, \bibinfo {author} {\bibfnamefont {E.}~\bibnamefont {{Platen}}},
  \bibinfo {author} {\bibfnamefont {B.}~\bibnamefont {{Beygu}}}, \bibinfo
  {author} {\bibfnamefont {J.~H.}\ \bibnamefont {{van Gorkom}}}, \bibinfo
  {author} {\bibfnamefont {J.~M.}\ \bibnamefont {{van der Hulst}}}, \bibinfo
  {author} {\bibfnamefont {M.~A.}\ \bibnamefont {{Arag{\'o}n-Calvo}}}, \bibinfo
  {author} {\bibfnamefont {P.~J.~E.}\ \bibnamefont {{Peebles}}}, \bibinfo
  {author} {\bibfnamefont {T.}~\bibnamefont {{Jarrett}}}, \bibinfo {author}
  {\bibfnamefont {G.}~\bibnamefont {{Rhee}}}, \bibinfo {author} {\bibfnamefont
  {K.}~\bibnamefont {{Kova{\v{c}}}}},\ and\ \bibinfo {author} {\bibfnamefont
  {C.~W.}\ \bibnamefont {{Yip}}},\ }\bibfield  {title} {\bibinfo {title} {{The
  Void Galaxy Survey}},\ }\href {https://doi.org/10.1007/978-3-642-20285-8_3}
  {\bibfield  {journal} {\bibinfo  {journal} {Astrophysics and Space Science
  Proceedings}\ }\textbf {\bibinfo {volume} {27}},\ \bibinfo {pages} {17}
  (\bibinfo {year} {2011})},\ \Eprint {https://arxiv.org/abs/1101.4187}
  {arXiv:1101.4187 [astro-ph.CO]} \BibitemShut {NoStop}%
\bibitem [{\citenamefont {{Pisani}}\ \emph {et~al.}(2019)\citenamefont
  {{Pisani}}, \citenamefont {{Massara}}, \citenamefont {{Spergel}},
  \citenamefont {{Alonso}}, \citenamefont {{Baker}}, \citenamefont {{Cai}},
  \citenamefont {{Cautun}}, \citenamefont {{Davies}}, \citenamefont
  {{Demchenko}}, \citenamefont {{Dor{\'e}}}, \citenamefont {{Goulding}},
  \citenamefont {{Habouzit}}, \citenamefont {{Hamaus}}, \citenamefont
  {{Hawken}}, \citenamefont {{Hirata}}, \citenamefont {{Ho}}, \citenamefont
  {{Jain}}, \citenamefont {{Kreisch}}, \citenamefont {{Marulli}}, \citenamefont
  {{Padilla}}, \citenamefont {{Pollina}}, \citenamefont {{Sahl{\'e}n}},
  \citenamefont {{Sheth}}, \citenamefont {{Somerville}}, \citenamefont
  {{Szapudi}}, \citenamefont {{van de Weygaert}}, \citenamefont
  {{Villaescusa-Navarro}}, \citenamefont {{Wandelt}},\ and\ \citenamefont
  {{Wang}}}]{whitepaper}%
  \BibitemOpen
  \bibfield  {author} {\bibinfo {author} {\bibfnamefont {A.}~\bibnamefont
  {{Pisani}}}, \bibinfo {author} {\bibfnamefont {E.}~\bibnamefont {{Massara}}},
  \bibinfo {author} {\bibfnamefont {D.~N.}\ \bibnamefont {{Spergel}}}, \bibinfo
  {author} {\bibfnamefont {D.}~\bibnamefont {{Alonso}}}, \bibinfo {author}
  {\bibfnamefont {T.}~\bibnamefont {{Baker}}}, \bibinfo {author} {\bibfnamefont
  {Y.-C.}\ \bibnamefont {{Cai}}}, \bibinfo {author} {\bibfnamefont
  {M.}~\bibnamefont {{Cautun}}}, \bibinfo {author} {\bibfnamefont
  {C.}~\bibnamefont {{Davies}}}, \bibinfo {author} {\bibfnamefont
  {V.}~\bibnamefont {{Demchenko}}}, \bibinfo {author} {\bibfnamefont
  {O.}~\bibnamefont {{Dor{\'e}}}}, \bibinfo {author} {\bibfnamefont
  {A.}~\bibnamefont {{Goulding}}}, \bibinfo {author} {\bibfnamefont
  {M.}~\bibnamefont {{Habouzit}}}, \bibinfo {author} {\bibfnamefont
  {N.}~\bibnamefont {{Hamaus}}}, \bibinfo {author} {\bibfnamefont
  {A.}~\bibnamefont {{Hawken}}}, \bibinfo {author} {\bibfnamefont {C.~M.}\
  \bibnamefont {{Hirata}}}, \bibinfo {author} {\bibfnamefont {S.}~\bibnamefont
  {{Ho}}}, \bibinfo {author} {\bibfnamefont {B.}~\bibnamefont {{Jain}}},
  \bibinfo {author} {\bibfnamefont {C.~D.}\ \bibnamefont {{Kreisch}}}, \bibinfo
  {author} {\bibfnamefont {F.}~\bibnamefont {{Marulli}}}, \bibinfo {author}
  {\bibfnamefont {N.}~\bibnamefont {{Padilla}}}, \bibinfo {author}
  {\bibfnamefont {G.}~\bibnamefont {{Pollina}}}, \bibinfo {author}
  {\bibfnamefont {M.}~\bibnamefont {{Sahl{\'e}n}}}, \bibinfo {author}
  {\bibfnamefont {R.~K.}\ \bibnamefont {{Sheth}}}, \bibinfo {author}
  {\bibfnamefont {R.}~\bibnamefont {{Somerville}}}, \bibinfo {author}
  {\bibfnamefont {I.}~\bibnamefont {{Szapudi}}}, \bibinfo {author}
  {\bibfnamefont {R.}~\bibnamefont {{van de Weygaert}}}, \bibinfo {author}
  {\bibfnamefont {F.}~\bibnamefont {{Villaescusa-Navarro}}}, \bibinfo {author}
  {\bibfnamefont {B.~D.}\ \bibnamefont {{Wandelt}}},\ and\ \bibinfo {author}
  {\bibfnamefont {Y.}~\bibnamefont {{Wang}}},\ }\bibfield  {title} {\bibinfo
  {title} {{Cosmic voids: a novel probe to shed light on our Universe}},\
  }\href@noop {} {\bibfield  {journal} {\bibinfo  {journal} {\baas}\ }\textbf
  {\bibinfo {volume} {51}},\ \bibinfo {eid} {40} (\bibinfo {year} {2019})},\
  \Eprint {https://arxiv.org/abs/1903.05161} {arXiv:1903.05161 [astro-ph.CO]}
  \BibitemShut {NoStop}%
\bibitem [{\citenamefont {{Hamaus}}\ \emph
  {et~al.}(2014{\natexlab{a}})\citenamefont {{Hamaus}}, \citenamefont
  {{Sutter}},\ and\ \citenamefont {{Wandelt}}}]{hamaus2014jun}%
  \BibitemOpen
  \bibfield  {author} {\bibinfo {author} {\bibfnamefont {N.}~\bibnamefont
  {{Hamaus}}}, \bibinfo {author} {\bibfnamefont {P.~M.}\ \bibnamefont
  {{Sutter}}},\ and\ \bibinfo {author} {\bibfnamefont {B.~D.}\ \bibnamefont
  {{Wandelt}}},\ }\bibfield  {title} {\bibinfo {title} {{Universal Density
  Profile for Cosmic Voids}},\ }\href
  {https://doi.org/10.1103/PhysRevLett.112.251302} {\bibfield  {journal}
  {\bibinfo  {journal} {\prl}\ }\textbf {\bibinfo {volume} {112}},\ \bibinfo
  {eid} {251302} (\bibinfo {year} {2014}{\natexlab{a}})},\ \Eprint
  {https://arxiv.org/abs/1403.5499} {arXiv:1403.5499 [astro-ph.CO]}
  \BibitemShut {NoStop}%
\bibitem [{\citenamefont {{Hamaus}}\ \emph
  {et~al.}(2014{\natexlab{b}})\citenamefont {{Hamaus}}, \citenamefont
  {{Sutter}}, \citenamefont {{Lavaux}},\ and\ \citenamefont {{Wand
  elt}}}]{hamaus2014dec}%
  \BibitemOpen
  \bibfield  {author} {\bibinfo {author} {\bibfnamefont {N.}~\bibnamefont
  {{Hamaus}}}, \bibinfo {author} {\bibfnamefont {P.~M.}\ \bibnamefont
  {{Sutter}}}, \bibinfo {author} {\bibfnamefont {G.}~\bibnamefont {{Lavaux}}},\
  and\ \bibinfo {author} {\bibfnamefont {B.~D.}\ \bibnamefont {{Wand elt}}},\
  }\bibfield  {title} {\bibinfo {title} {{Testing cosmic geometry without
  dynamic distortions using voids}},\ }\href
  {https://doi.org/10.1088/1475-7516/2014/12/013} {\bibfield  {journal}
  {\bibinfo  {journal} {\jcap}\ }\textbf {\bibinfo {volume} {2014}},\ \bibinfo
  {eid} {013} (\bibinfo {year} {2014}{\natexlab{b}})},\ \Eprint
  {https://arxiv.org/abs/1409.3580} {arXiv:1409.3580 [astro-ph.CO]}
  \BibitemShut {NoStop}%
\bibitem [{\citenamefont {Tanabashi}\ \emph {et~al.}(2018)\citenamefont
  {Tanabashi}, \citenamefont {Hagiwara}, \citenamefont {Hikasa}, \citenamefont
  {Nakamura}, \citenamefont {Sumino}, \citenamefont {Takahashi}, \citenamefont
  {Tanaka}, \citenamefont {Agashe}, \citenamefont {Aielli}, \citenamefont
  {Amsler}, \citenamefont {Antonelli}, \citenamefont {Asner}, \citenamefont
  {Baer}, \citenamefont {Banerjee}, \citenamefont {Barnett}, \citenamefont
  {Basaglia}, \citenamefont {Bauer}, \citenamefont {Beatty}, \citenamefont
  {Belousov}, \citenamefont {Beringer}, \citenamefont {Bethke}, \citenamefont
  {Bettini}, \citenamefont {Bichsel}, \citenamefont {Biebel}, \citenamefont
  {Black}, \citenamefont {Blucher}, \citenamefont {Buchmuller},\ and\
  \citenamefont {Burkert}}]{pdg}%
  \BibitemOpen
  \bibfield  {author} {\bibinfo {author} {\bibfnamefont {M.}~\bibnamefont
  {Tanabashi}}, \bibinfo {author} {\bibfnamefont {K.}~\bibnamefont {Hagiwara}},
  \bibinfo {author} {\bibfnamefont {K.}~\bibnamefont {Hikasa}}, \bibinfo
  {author} {\bibfnamefont {K.}~\bibnamefont {Nakamura}}, \bibinfo {author}
  {\bibfnamefont {Y.}~\bibnamefont {Sumino}}, \bibinfo {author} {\bibfnamefont
  {F.}~\bibnamefont {Takahashi}}, \bibinfo {author} {\bibfnamefont
  {J.}~\bibnamefont {Tanaka}}, \bibinfo {author} {\bibfnamefont
  {K.}~\bibnamefont {Agashe}}, \bibinfo {author} {\bibfnamefont
  {G.}~\bibnamefont {Aielli}}, \bibinfo {author} {\bibfnamefont
  {C.}~\bibnamefont {Amsler}}, \bibinfo {author} {\bibfnamefont
  {M.}~\bibnamefont {Antonelli}}, \bibinfo {author} {\bibfnamefont {D.~M.}\
  \bibnamefont {Asner}}, \bibinfo {author} {\bibfnamefont {H.}~\bibnamefont
  {Baer}}, \bibinfo {author} {\bibfnamefont {S.}~\bibnamefont {Banerjee}},
  \bibinfo {author} {\bibfnamefont {R.~M.}\ \bibnamefont {Barnett}}, \bibinfo
  {author} {\bibfnamefont {T.}~\bibnamefont {Basaglia}}, \bibinfo {author}
  {\bibfnamefont {C.~W.}\ \bibnamefont {Bauer}}, \bibinfo {author}
  {\bibfnamefont {J.~J.}\ \bibnamefont {Beatty}}, \bibinfo {author}
  {\bibfnamefont {V.~I.}\ \bibnamefont {Belousov}}, \bibinfo {author}
  {\bibfnamefont {J.}~\bibnamefont {Beringer}}, \bibinfo {author}
  {\bibfnamefont {S.}~\bibnamefont {Bethke}}, \bibinfo {author} {\bibfnamefont
  {A.}~\bibnamefont {Bettini}}, \bibinfo {author} {\bibfnamefont
  {H.}~\bibnamefont {Bichsel}}, \bibinfo {author} {\bibfnamefont
  {O.}~\bibnamefont {Biebel}}, \bibinfo {author} {\bibfnamefont {K.~M.}\
  \bibnamefont {Black}}, \bibinfo {author} {\bibfnamefont {E.}~\bibnamefont
  {Blucher}}, \bibinfo {author} {\bibfnamefont {O.}~\bibnamefont
  {Buchmuller}},\ and\ \bibinfo {author} {\bibfnamefont {V.}~\bibnamefont
  {Burkert}} (\bibinfo {collaboration} {Particle Data Group}),\ }\bibfield
  {title} {\bibinfo {title} {Review of particle physics},\ }\href
  {https://doi.org/10.1103/PhysRevD.98.030001} {\bibfield  {journal} {\bibinfo
  {journal} {Phys. Rev. D}\ }\textbf {\bibinfo {volume} {98}},\ \bibinfo
  {pages} {030001} (\bibinfo {year} {2018})}\BibitemShut {NoStop}%
\bibitem [{\citenamefont {Lesgourgues}\ and\ \citenamefont
  {Pastor}(2012)}]{lesgourgues2012}%
  \BibitemOpen
  \bibfield  {author} {\bibinfo {author} {\bibfnamefont {J.}~\bibnamefont
  {Lesgourgues}}\ and\ \bibinfo {author} {\bibfnamefont {S.}~\bibnamefont
  {Pastor}},\ }\bibfield  {title} {\bibinfo {title} {{Neutrino mass from
  Cosmology}},\ }\href {https://doi.org/10.1155/2012/608515} {\bibfield
  {journal} {\bibinfo  {journal} {Adv. High Energy Phys.}\ }\textbf {\bibinfo
  {volume} {2012}},\ \bibinfo {pages} {608515} (\bibinfo {year} {2012})},\
  \Eprint {https://arxiv.org/abs/1212.6154} {arXiv:1212.6154 [hep-ph]}
  \BibitemShut {NoStop}%
%%CITATION = ARXIV:1212.6154;%%
\bibitem [{\citenamefont {Liu}\ \emph {et~al.}(2018)\citenamefont {Liu},
  \citenamefont {Bird}, \citenamefont {Matilla}, \citenamefont {Hill},
  \citenamefont {Haiman}, \citenamefont {Madhavacheril}, \citenamefont
  {Petri},\ and\ \citenamefont {Spergel}}]{jia2018}%
  \BibitemOpen
  \bibfield  {author} {\bibinfo {author} {\bibfnamefont {J.}~\bibnamefont
  {Liu}}, \bibinfo {author} {\bibfnamefont {S.}~\bibnamefont {Bird}}, \bibinfo
  {author} {\bibfnamefont {J.~M.~Z.}\ \bibnamefont {Matilla}}, \bibinfo
  {author} {\bibfnamefont {J.~C.}\ \bibnamefont {Hill}}, \bibinfo {author}
  {\bibfnamefont {Z.}~\bibnamefont {Haiman}}, \bibinfo {author} {\bibfnamefont
  {M.~S.}\ \bibnamefont {Madhavacheril}}, \bibinfo {author} {\bibfnamefont
  {A.}~\bibnamefont {Petri}},\ and\ \bibinfo {author} {\bibfnamefont {D.~N.}\
  \bibnamefont {Spergel}},\ }\bibfield  {title} {\bibinfo {title}
  {{MassiveNuS}: cosmological massive neutrino simulations},\ }\href
  {https://doi.org/10.1088/1475-7516/2018/03/049} {\bibfield  {journal}
  {\bibinfo  {journal} {Journal of Cosmology and Astroparticle Physics}\
  }\textbf {\bibinfo {volume} {2018}}\bibinfo  {number} { (03)},\ \bibinfo
  {pages} {049}}\BibitemShut {NoStop}%
\bibitem [{\citenamefont {{Sutter}}\ \emph
  {et~al.}(2015{\natexlab{b}})\citenamefont {{Sutter}}, \citenamefont
  {{Lavaux}}, \citenamefont {{Hamaus}}, \citenamefont {{Pisani}}, \citenamefont
  {{Wandelt}}, \citenamefont {{Warren}}, \citenamefont {{Villaescusa-Navarro}},
  \citenamefont {{Zivick}}, \citenamefont {{Mao}},\ and\ \citenamefont
  {{Thompson}}}]{vide}%
  \BibitemOpen
\bibfield  {number} {  }\bibfield  {author} {\bibinfo {author} {\bibfnamefont
  {P.~M.}\ \bibnamefont {{Sutter}}}, \bibinfo {author} {\bibfnamefont
  {G.}~\bibnamefont {{Lavaux}}}, \bibinfo {author} {\bibfnamefont
  {N.}~\bibnamefont {{Hamaus}}}, \bibinfo {author} {\bibfnamefont
  {A.}~\bibnamefont {{Pisani}}}, \bibinfo {author} {\bibfnamefont {B.~D.}\
  \bibnamefont {{Wandelt}}}, \bibinfo {author} {\bibfnamefont {M.}~\bibnamefont
  {{Warren}}}, \bibinfo {author} {\bibfnamefont {F.}~\bibnamefont
  {{Villaescusa-Navarro}}}, \bibinfo {author} {\bibfnamefont {P.}~\bibnamefont
  {{Zivick}}}, \bibinfo {author} {\bibfnamefont {Q.}~\bibnamefont {{Mao}}},\
  and\ \bibinfo {author} {\bibfnamefont {B.~B.}\ \bibnamefont {{Thompson}}},\
  }\bibfield  {title} {\bibinfo {title} {{VIDE: The Void IDentification and
  Examination toolkit}},\ }\href {https://doi.org/10.1016/j.ascom.2014.10.002}
  {\bibfield  {journal} {\bibinfo  {journal} {Astronomy and Computing}\
  }\textbf {\bibinfo {volume} {9}},\ \bibinfo {pages} {1} (\bibinfo {year}
  {2015}{\natexlab{b}})},\ \Eprint {https://arxiv.org/abs/1406.1191}
  {arXiv:1406.1191 [astro-ph.CO]} \BibitemShut {NoStop}%
\bibitem [{\citenamefont {{Springel}}(2005)}]{gadget2}%
  \BibitemOpen
  \bibfield  {author} {\bibinfo {author} {\bibfnamefont {V.}~\bibnamefont
  {{Springel}}},\ }\bibfield  {title} {\bibinfo {title} {{The cosmological
  simulation code GADGET-2}},\ }\href
  {https://doi.org/10.1111/j.1365-2966.2005.09655.x} {\bibfield  {journal}
  {\bibinfo  {journal} {\mnras}\ }\textbf {\bibinfo {volume} {364}},\ \bibinfo
  {pages} {1105} (\bibinfo {year} {2005})},\ \Eprint
  {https://arxiv.org/abs/astro-ph/0505010} {arXiv:astro-ph/0505010 [astro-ph]}
  \BibitemShut {NoStop}%
\bibitem [{\citenamefont {{Press}}\ and\ \citenamefont
  {{Schechter}}(1974)}]{HMFpress}%
  \BibitemOpen
  \bibfield  {author} {\bibinfo {author} {\bibfnamefont {W.~H.}\ \bibnamefont
  {{Press}}}\ and\ \bibinfo {author} {\bibfnamefont {P.}~\bibnamefont
  {{Schechter}}},\ }\bibfield  {title} {\bibinfo {title} {{Formation of
  Galaxies and Clusters of Galaxies by Self-Similar Gravitational
  Condensation}},\ }\href {https://doi.org/10.1086/152650} {\bibfield
  {journal} {\bibinfo  {journal} {\apj}\ }\textbf {\bibinfo {volume} {187}},\
  \bibinfo {pages} {425} (\bibinfo {year} {1974})}\BibitemShut {NoStop}%
\bibitem [{\citenamefont {{Bond}}\ \emph {et~al.}(1991)\citenamefont {{Bond}},
  \citenamefont {{Cole}}, \citenamefont {{Efstathiou}},\ and\ \citenamefont
  {{Kaiser}}}]{HMFbond}%
  \BibitemOpen
  \bibfield  {author} {\bibinfo {author} {\bibfnamefont {J.~R.}\ \bibnamefont
  {{Bond}}}, \bibinfo {author} {\bibfnamefont {S.}~\bibnamefont {{Cole}}},
  \bibinfo {author} {\bibfnamefont {G.}~\bibnamefont {{Efstathiou}}},\ and\
  \bibinfo {author} {\bibfnamefont {N.}~\bibnamefont {{Kaiser}}},\ }\bibfield
  {title} {\bibinfo {title} {{Excursion set mass functions for hierarchical
  Gaussian fluctuations}},\ }\href {https://doi.org/10.1086/170520} {\bibfield
  {journal} {\bibinfo  {journal} {\apj}\ }\textbf {\bibinfo {volume} {379}},\
  \bibinfo {pages} {440} (\bibinfo {year} {1991})}\BibitemShut {NoStop}%
\bibitem [{\citenamefont {Bartelmann}\ and\ \citenamefont
  {Schneider}(2001)}]{lensing}%
  \BibitemOpen
  \bibfield  {author} {\bibinfo {author} {\bibfnamefont {M.}~\bibnamefont
  {Bartelmann}}\ and\ \bibinfo {author} {\bibfnamefont {P.}~\bibnamefont
  {Schneider}},\ }\bibfield  {title} {\bibinfo {title} {{Weak gravitational
  lensing}},\ }\href {https://doi.org/10.1016/S0370-1573(00)00082-X} {\bibfield
   {journal} {\bibinfo  {journal} {Phys. Rept.}\ }\textbf {\bibinfo {volume}
  {340}},\ \bibinfo {pages} {291} (\bibinfo {year} {2001})},\ \Eprint
  {https://arxiv.org/abs/astro-ph/9912508} {arXiv:astro-ph/9912508 [astro-ph]}
  \BibitemShut {NoStop}%
%%CITATION = ASTRO-PH/9912508;%%
\bibitem [{\citenamefont {{Mandelbaum}}\ \emph {et~al.}(2006)\citenamefont
  {{Mandelbaum}}, \citenamefont {{Seljak}}, \citenamefont {{Cool}},
  \citenamefont {{Blanton}}, \citenamefont {{Hirata}},\ and\ \citenamefont
  {{Brinkmann}}}]{stacked}%
  \BibitemOpen
  \bibfield  {author} {\bibinfo {author} {\bibfnamefont {R.}~\bibnamefont
  {{Mandelbaum}}}, \bibinfo {author} {\bibfnamefont {U.}~\bibnamefont
  {{Seljak}}}, \bibinfo {author} {\bibfnamefont {R.~J.}\ \bibnamefont
  {{Cool}}}, \bibinfo {author} {\bibfnamefont {M.}~\bibnamefont {{Blanton}}},
  \bibinfo {author} {\bibfnamefont {C.~M.}\ \bibnamefont {{Hirata}}},\ and\
  \bibinfo {author} {\bibfnamefont {J.}~\bibnamefont {{Brinkmann}}},\
  }\bibfield  {title} {\bibinfo {title} {{Density profiles of galaxy groups and
  clusters from SDSS galaxy-galaxy weak lensing}},\ }\href
  {https://doi.org/10.1111/j.1365-2966.2006.10906.x} {\bibfield  {journal}
  {\bibinfo  {journal} {\mnras}\ }\textbf {\bibinfo {volume} {372}},\ \bibinfo
  {pages} {758} (\bibinfo {year} {2006})},\ \Eprint
  {https://arxiv.org/abs/astro-ph/0605476} {arXiv:astro-ph/0605476 [astro-ph]}
  \BibitemShut {NoStop}%
\bibitem [{\citenamefont {{Huang}}\ \emph {et~al.}(2018)\citenamefont
  {{Huang}}, \citenamefont {{Leauthaud}}, \citenamefont {{Hearin}},
  \citenamefont {{Behroozi}}, \citenamefont {{Bradshaw}}, \citenamefont
  {{Ardila}}, \citenamefont {{Speagle}}, \citenamefont {{Tenenti}},
  \citenamefont {{Bundy}}, \citenamefont {{Greene}}, \citenamefont {{Sifon}},\
  and\ \citenamefont {{Bahcall}}}]{stacked2}%
  \BibitemOpen
  \bibfield  {author} {\bibinfo {author} {\bibfnamefont {S.}~\bibnamefont
  {{Huang}}}, \bibinfo {author} {\bibfnamefont {A.}~\bibnamefont
  {{Leauthaud}}}, \bibinfo {author} {\bibfnamefont {A.}~\bibnamefont
  {{Hearin}}}, \bibinfo {author} {\bibfnamefont {P.}~\bibnamefont
  {{Behroozi}}}, \bibinfo {author} {\bibfnamefont {C.}~\bibnamefont
  {{Bradshaw}}}, \bibinfo {author} {\bibfnamefont {F.}~\bibnamefont
  {{Ardila}}}, \bibinfo {author} {\bibfnamefont {J.}~\bibnamefont {{Speagle}}},
  \bibinfo {author} {\bibfnamefont {A.}~\bibnamefont {{Tenenti}}}, \bibinfo
  {author} {\bibfnamefont {K.}~\bibnamefont {{Bundy}}}, \bibinfo {author}
  {\bibfnamefont {J.}~\bibnamefont {{Greene}}}, \bibinfo {author}
  {\bibfnamefont {C.}~\bibnamefont {{Sifon}}},\ and\ \bibinfo {author}
  {\bibfnamefont {N.}~\bibnamefont {{Bahcall}}},\ }\bibfield  {title} {\bibinfo
  {title} {{Weak Lensing Reveals a Tight Connection Between Dark Matter Halo
  Mass and the Distribution of Stellar Mass in Massive Galaxies}},\ }\href@noop
  {} {\bibfield  {journal} {\bibinfo  {journal} {arXiv e-prints}\ ,\ \bibinfo
  {eid} {arXiv:1811.01139}} (\bibinfo {year} {2018})},\ \Eprint
  {https://arxiv.org/abs/1811.01139} {arXiv:1811.01139 [astro-ph.GA]}
  \BibitemShut {NoStop}%
\bibitem [{\citenamefont {{Paillas}}\ \emph {et~al.}(2017)\citenamefont
  {{Paillas}}, \citenamefont {{Lagos}}, \citenamefont {{Padilla}},
  \citenamefont {{Tissera}}, \citenamefont {{Helly}},\ and\ \citenamefont
  {{Schaller}}}]{Paillas2017}%
  \BibitemOpen
  \bibfield  {author} {\bibinfo {author} {\bibfnamefont {E.}~\bibnamefont
  {{Paillas}}}, \bibinfo {author} {\bibfnamefont {C.~D.~P.}\ \bibnamefont
  {{Lagos}}}, \bibinfo {author} {\bibfnamefont {N.}~\bibnamefont {{Padilla}}},
  \bibinfo {author} {\bibfnamefont {P.}~\bibnamefont {{Tissera}}}, \bibinfo
  {author} {\bibfnamefont {J.}~\bibnamefont {{Helly}}},\ and\ \bibinfo {author}
  {\bibfnamefont {M.}~\bibnamefont {{Schaller}}},\ }\bibfield  {title}
  {\bibinfo {title} {{Baryon effects on void statistics in the EAGLE
  simulation}},\ }\href {https://doi.org/10.1093/mnras/stx1514} {\bibfield
  {journal} {\bibinfo  {journal} {\mnras}\ }\textbf {\bibinfo {volume} {470}},\
  \bibinfo {pages} {4434} (\bibinfo {year} {2017})},\ \Eprint
  {https://arxiv.org/abs/1609.00101} {arXiv:1609.00101 [astro-ph.CO]}
  \BibitemShut {NoStop}%
\end{thebibliography}%
\end{document}